\documentclass[12pt,preprint]{aastex}
\usepackage{multirow}
\usepackage{amsmath}
\usepackage{longtable}
\usepackage{rotating}
\usepackage{booktabs}
\usepackage{graphicx}
\usepackage{ulem}

\newcommand{\mj}{M_{\rm J}}
\newcommand{\rp}{r_{\rm p}}

\newcommand{\mplanet}{M_{\rm p}}

\begin{document}
\title{The Eccentric Cavity, Triple Rings, Two-Armed Spirals, and Double Clumps of the MWC 758 Disk}

\author{Ruobing Dong\altaffilmark{1,2}, Sheng-yuan Liu\altaffilmark{2}, Josh Eisner\altaffilmark{1}, Sean Andrews\altaffilmark{3}, Jeffrey Fung\altaffilmark{4}, Zhaohuan Zhu\altaffilmark{5}, Eugene Chiang\altaffilmark{4}, Jun Hashimoto\altaffilmark{6}, Hauyu Baobab Liu\altaffilmark{7},
Simon Casassus\altaffilmark{8}, Thomas Esposito\altaffilmark{4}, Yasuhiro Hasegawa\altaffilmark{9}, Takayuki Muto\altaffilmark{10}, Yaroslav Pavlyuchenkov\altaffilmark{11}, David Wilner\altaffilmark{3}, 
Eiji Akiyama\altaffilmark{12}, Motohide Tamura\altaffilmark{5,13}, and John Wisniewski\altaffilmark{14}}

\altaffiltext{1}{Steward Observatory, University of Arizona, AZ 85721, USA}
\altaffiltext{2}{Institute of Astronomy and Astrophysics, Academia Sinica, Taipei 10617, Taiwan}
\altaffiltext{3}{Harvard-Smithsonian Center for Astrophysics, 60 Garden Street, Cambridge, MA 02138, USA}
\altaffiltext{4}{Department of Astronomy, University of California at Berkeley, Campbell Hall, Berkeley, CA 94720-3411, USA}
\altaffiltext{5}{Department of Physics and Astronomy, University of Nevada, Las Vegas, 4505 South Maryland Parkway, Las Vegas, NV 89154, USA}
\altaffiltext{6}{Astrobiology Center of NINS, 2-21-1, Osawa, Mitaka, Tokyo 181-8588, Japan}
\altaffiltext{7}{European Southern Observatory (ESO), Karl-Schwarzschild-Str. 2, D-85748 Garching, Germany}
\altaffiltext{8}{Departamento de Astronomia, Universidad de Chile, Casilla 36-D, Santiago, Chile}
\altaffiltext{9}{Jet Propulsion Laboratory, California Institute of Technology, Pasadena, CA 91109, USA}
\altaffiltext{10}{Division of Liberal Arts, Kogakuin University, 1-24-2 Nishi-Shinjuku, Shinjuku-ku, Tokyo 163-8677, Japan}
\altaffiltext{11}{Institute of Astronomy, Russian Academy of Sciences, Moscow, Russia}
\altaffiltext{12}{National Astronomical Observatory of Japan, 2-21-1 Osawa, Mitaka, Tokyo 181-8588, Japan}
\altaffiltext{13}{Department of Astronomy, The University of Tokyo, 7-3-1, Hongo, Bunkyo-ku, Tokyo 113-0033, Japan}
\altaffiltext{14}{H. L. Dodge Department of Physics \& Astronomy, University of Oklahoma, 440 W Brooks St Norman, OK 73019, USA}

\clearpage

\begin{abstract}

Spatially resolved structures in protoplanetary disks hint at unseen planets. Previous imaging observations of the transitional disk around MWC 758 revealed an inner cavity, a ring-like outer disk, emission clumps, and spiral arms, all possibly generated by companions. We present ALMA dust continuum observations of MWC 758 at 0.87 millimeter (mm) wavelength with 43$\times$39 mas angular resolution (6.9$\times$6.2 AU) and 20 $\mu$Jy beam$^{-1}$ rms. The central sub-mm emission cavity is revealed to be eccentric; once deprojected, its outer edge can be well-fitted by an ellipse with an eccentricity of 0.1 and one focus on the star. The broad ring-like outer disk is resolved into three narrow rings with two gaps in between. The outer two rings tentatively show the same eccentricity and orientation as the innermost ring bounding the inner cavity. The two previously known dust emission clumps are resolved in both the radial and azimuthal directions, with radial widths equal to $\sim$4$\times$ the local scale height. Only one of the two spiral arms previously imaged in near-infrared (NIR) scattered light is revealed in ALMA dust emission, at a slightly larger stellocentric distance owing to projection effects. We also submit evidence of disk truncation at $\sim$100 AU based on comparing NIR imaging observations with models. The spirals, the north clump, and the truncated disk edge are all broadly consistent with the presence of one companion exterior to the spirals at roughly 100 AU.

\end{abstract}



\section{Introduction}\label{sec:intro}

Planets form in protoplanetary disks surrounding newborn stars typically one to a few million years old. Forming planets perturb the disks via gravitational interactions, and may produce large scale structures such as spiral density waves and gaps \citep[e.g.,][]{goldreich80, lin93, bryden99}. Such structures may have been detected in optical to near-infrared (NIR) scattered light imaging (e.g., spirals: \citealt{muto12, stolker17, uyama18, canovas18}; gaps: \citealt{debes13, pinilla15j1604, pohl17hd169142}) and (sub-)millimeter (mm) to centimeter (cm) interferometric observations of disks (e.g., spirals: \citealt{perez16}; gaps: \citealt{canovas15sz91, isella16hd163296, dong17j1604, dipierro18}). Disk gaps, spiral arms, and other structures may help infer the presence of (unseen) planets forming in disks, and constrain fundamental parameters of these planets, such as mass, location, and orbit.

Here we target a young star+disk system, MWC 758, with previously-imaged structures suggestive of carving by planetary-mass companions. MWC 758 is a 3.5$\pm$2 Myr \citep{meeus12} Herbig Ae star located at a distance of 160$\pm$2 pc \citep{gaia18}. It is surrounded by a protoplanetary disk in Keplerian rotation \citep{isella10}. The disk has a large cavity $\sim$50~AU in radius in sub-mm continuum emission \citep{andrews11}. It also has a set of near-symmetric two-arm spirals at $\sim$30$-$75 AU in NIR scattered light \citep{grady13, benisty15}, and two emission clumps at tens of AU in sub-mm to cm continuum emission \citep{marino15mwc758, casassus18mwc758}. \citet{boehler18} presented ALMA cycle 3 observations of MWC 758 in both 0.88 mm continuum emission and in $^{13}$CO/C$^{18}$O $J$=3-2 emission with 0\farcs1-0\farcs2 angular resolution. The known $\sim$mm dust ring was resolved into a double-ring, and a compact emission source centered on the star was identified. The spiral arms were seen in $^{13}$CO emission. 

Generally speaking, cavities, spiral arms, and emission clumps may reflect gaps, density waves, and vortices dynamically induced by disk-embedded planets \citep[e.g.,][]{zhu11, lyra13, zhu14stone, bae16sao206462, hammer17}. Specifically, \citet{dong15spiralarm} showed that the MWC 758 disk's two arms can be quantitatively explained as the primary and secondary spiral shocks driven by one multi-$\mj$ mass companion exterior to the arms. Very recently, \citet{ren18} measured the pattern speed of the arms using multi-epoch scattered light observations, and concluded that the best-fit pattern speed corresponds to the Keplerian speed at $r\sim 90$ AU from the star. This is consistent with the hypothesis that they are excited by a companion at that radius.

In this paper, we present ALMA Cycle 5 continuum emission observations of the MWC 758 disk at 0.87 mm with an angular resolution of 43$\times$39 mas (6.8$\times$6.2 AU), comparable to the angular resolution in NIR direct imaging (0\farcs04 at $H$-band) and roughly four times better than the highest angular resolution achieved in previous $\sim$mm observations of this target. Nearly all major known disk features are confirmed and further resolved. The main new discoveries include the ellipticity of the central cavity, a triple-ring structure most noticeable toward the west, and the sub-mm continuum emission counterpart of the southern spiral arm imaged in scattered light. In addition, both emission clumps are now resolved in the radial direction for the first time.


\section{ALMA Observations and Calibration}\label{sec:ALMA}

Our observations, including three execution blocks (EBs) toward MWC 758, were carried out in Cycle 5 by ALMA under the project 2017.1.00492.S. The first EB was conducted on 2017 November 12 with 48 antennas in the array configuration C43-8. The other two EBs were conducted on 2017 November 25 with 48 and 49 antennas in the same C43-8 array configuration. The combined dataset has baselines ranging between 92~m and 11.84~km, which correspond to angular scales between $\sim$1\farcs8 and 0\farcs009. The total integration time on-source is about 125 minutes. Typical precipitable water vapor during the on-source period ranges between 0.55~mm and 0.8~mm.

Four spectral windows centered at 336.495~GHz, 338.432~GHz, 348.495~GHz, and 350.495~GHz, each with an effective bandwidth of 1.78125 GHz and 128 spectral channels under the Time Division Mode (TDM), were employed to maximize the continuum sensitivity. During all three executions, J0510+1800 served as the pointing, bandpass, amplitude, and check calibrator while J0521+2112 was used as the phase calibrator. The flux scale was calibrated against J0510+1800.  which has a spectral index of -0.375 and fluxes set as 1.331~Jy, 1.445~Jy, and 1.445~Jy at 348.495~GHz for the three EBs, respectively.

Data were first processed and calibrated through the ALMA pipeline calibration procedures under the Common Astronomy Software Applications \citep[CASA;][]{mcmullin07}. Imaging of the continuum emission is subsequently achieved by using all the four TDM spectral windows with potential spectral contamination inspected. We used the Briggs weighting scheme with a robust parameter of 0.5 and an additional $uv$-tapering (uvtaper parameter = [80000k$\lambda$, 4500k$\lambda$, -1$^\circ$] in the {\tt tclean} task) for forming a relatively circular beam and improving the signal-to-noise ratio. This results in a synthesized beam size of 43$\times$39 mas (6.8$\times$6.2 AU) at PA=-4.3$^\circ$, and a rms noise level of 20 $\mu$Jy beam$^{-1}$. The continuum emission is detected with a peak signal-to-noise ratio of 80. The reduced data cube is available as ApJ online supplemental material. Another version of the image synthesized using the natural weighting scheme (not shown in the paper) can also be found in the online supplemental material.


\section{Main Features}\label{sec:results}

Figure~\ref{fig:image} shows the synthesized image of the continuum emission from MWC 758 at 343.5 GHz (0.87~mm). The disk is detected at $\geq$3$\sigma$ out to $\sim$0\farcs64 (102 AU; see the $3\sigma$ contour in panel ($b$)). This high resolution map reveals rich features, including a central cavity, a broad outer disk composed of three narrow rings (inner, middle, and outer ring) and two narrow gaps in between, a north clump at the outer edge of the disk, a south clump at the outer edge of the cavity, a central point source, and a spiral arm in the south. Table~\ref{tab:measurement} lists the location, size, and flux measurements of some of the features. The total flux density integrated over a circle of a diameter of 1\farcs3 in size encompassing the disk extent is 180~mJy, comparable to the values of 180 mJy at 340~GHz measured with the Submillimeter Array by \cite{andrews11} and 205 mJy at 337~GHz with ALMA by \cite{marino15mwc758}. The differences at $\sim$10\% level may arise from flux calibration uncertainties or missing short $uv$ sampling. In the latter case, the missing flux would correspond to a smooth and low surface brightness structure roughly $\sim$30 $\mu$Jy beam$^{-1}$ or 1.5$\sigma$ in brightness. The resulting effects in the measurements of amplitude peaks and contrasts are small.

\subsection{The Central Point Source and the Non-detection of CPDs}\label{sec:results_pointsources}

The central point source (inset, Figure~\ref{fig:image}$b$) is detected at the 9$\sigma$ level with a peak flux of 0.17 mJy beam$^{-1}$. The resolution constrains the point source size to be $<$ 3 AU in radius. Given the incomplete short spacing coverage, there is a residual negative bowl within the central cavity at a 0.04~mJy beam$^{-1}$ level ($\sim$2$\sigma$). When fitted with a circular Gaussian with noises and the negative floor inside the cavity factored in, the actual integrated flux density of the central source is around 0.18~mJy. Same as \citet{boehler18}, in which the point source was also detected around the location of the star, we assume it originates from a small circumstellar structure previously detected in infrared interferometric observations \citep{eisner04}. A two-year baseline established by the \citet{boehler18} observations and our observations rules out the source being a background (not-moving) object based on MWC 758's propermotion \citep[27 mas yr$^{-1}$;][]{gaia18}. Combining with the total flux density of the central source measured at $\nu$=33GHz by the VLA \citep[$F$(33GHz)=67 $\mu$Jy,][]{marino15mwc758}, we derive a spectral index $\alpha$=$\log{\left((F(\nu1)/F(\nu2)\right)}/\log{(\nu1/\nu2)}$=0.4 between 343.5 and 33 GHz, consistent with the $\alpha$ derived in between 33 GHz and 15 GHz \citep[0.36;][]{marino15mwc758}. This value is significantly flatter than $\alpha$=2, expected if the source is optically thick, or $\alpha$=2+$\beta$, where $\beta$ is the dust opacity index (usually a positive number; \citealt{draine06}), expected if the source is optically thin. Such a low value indicates contribution (or dominance) of free-free emission from ionized gas (e.g., disk wind) very close to the star \citep[e.g.,][]{liu17fuori}. Alternatively, it may indicate self-absorbed dust emission in the dense $\sim$AU-scale circumstellar region \citep{li17}.

We note that the fitted Gaussian position of this source is offset by $\sim$7 mas to the south-southeast from the \citet[the propermotion of the star has been accounted for]{gaia18} stellar location (same as the phase center). This may be due to astrometry errors in phase referencing or position uncertainties from phase noise. The astrometry error based on the check source J0510+1800 is on average 2~mas. The position uncertainty due to phase noise is at a level of 2~mas, given the signal-to-noise ratio and the beam size (i.e., 0.45$\times$beam size / SNR; \citealt{reid88, wright90}). There remains the possibility of a true sub-AU scale asymmetry in the inner disk to be examined in the future.

Other than the central source, we do not detect any compact point source at the $3\sigma$ level that may be associated with circumplanetary disks (CPD) around forming planets. Hydro simulations have shown that the radius of a circumcompanion disk is $\sim$1/3 of the companion's Hill radius \citep{martin11}. At $r\leq100$AU, the diameter of such a disk is smaller than 10AU, or twice the beam size of our ALMA observation, for any object with $\mplanet\leq13\mj$. Considering that sub-mm continuum emission from such a disk is probably centrally peaked and compact \citep[e.g.,][]{wu17}, we do not expect to significantly resolve such a disk at $r\leq100$AU if detected (see also simulated ALMA images of CPDs in \citealt{szulagyi18}). Inside the cavity and outside the $3\sigma$ contour at $r$$\sim$0\farcs64 in Figure~\ref{fig:image}$b$ where the emission reaches the noise floor, the $3\sigma$ upper limit on the flux density of unresolved point sources is 60$\mu$Jy. Within the broad outer disk the detection limit is higher due to background disk emission.

\subsection{The Cavity and the Outer Disk}\label{sec:results_cavity}

\begin{figure}
\begin{center}
\includegraphics[trim=0 0 0 0, clip,width=0.45\textwidth,angle=0]{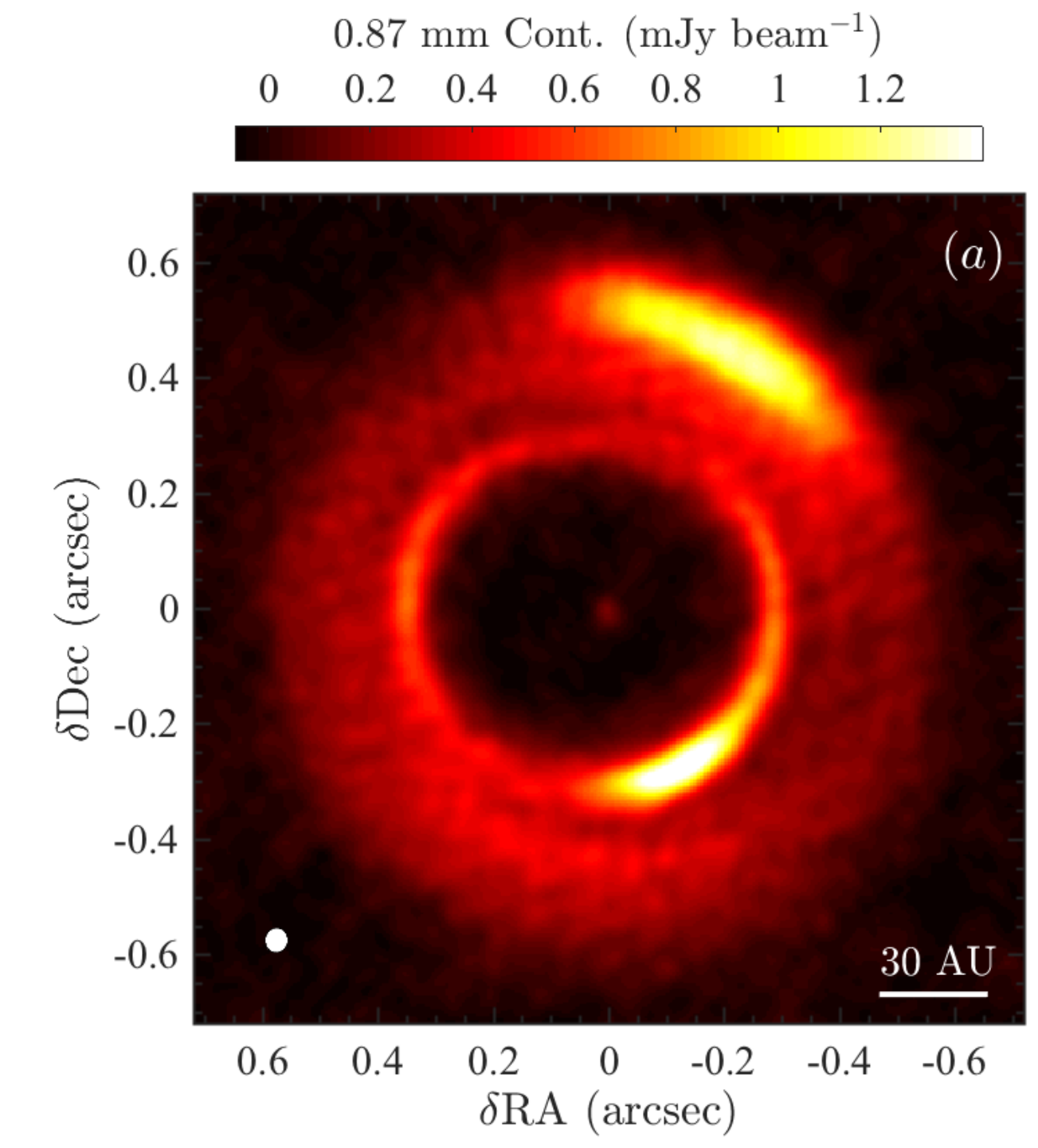}
\includegraphics[trim=0 0 0 0, clip,width=0.45\textwidth,angle=0]{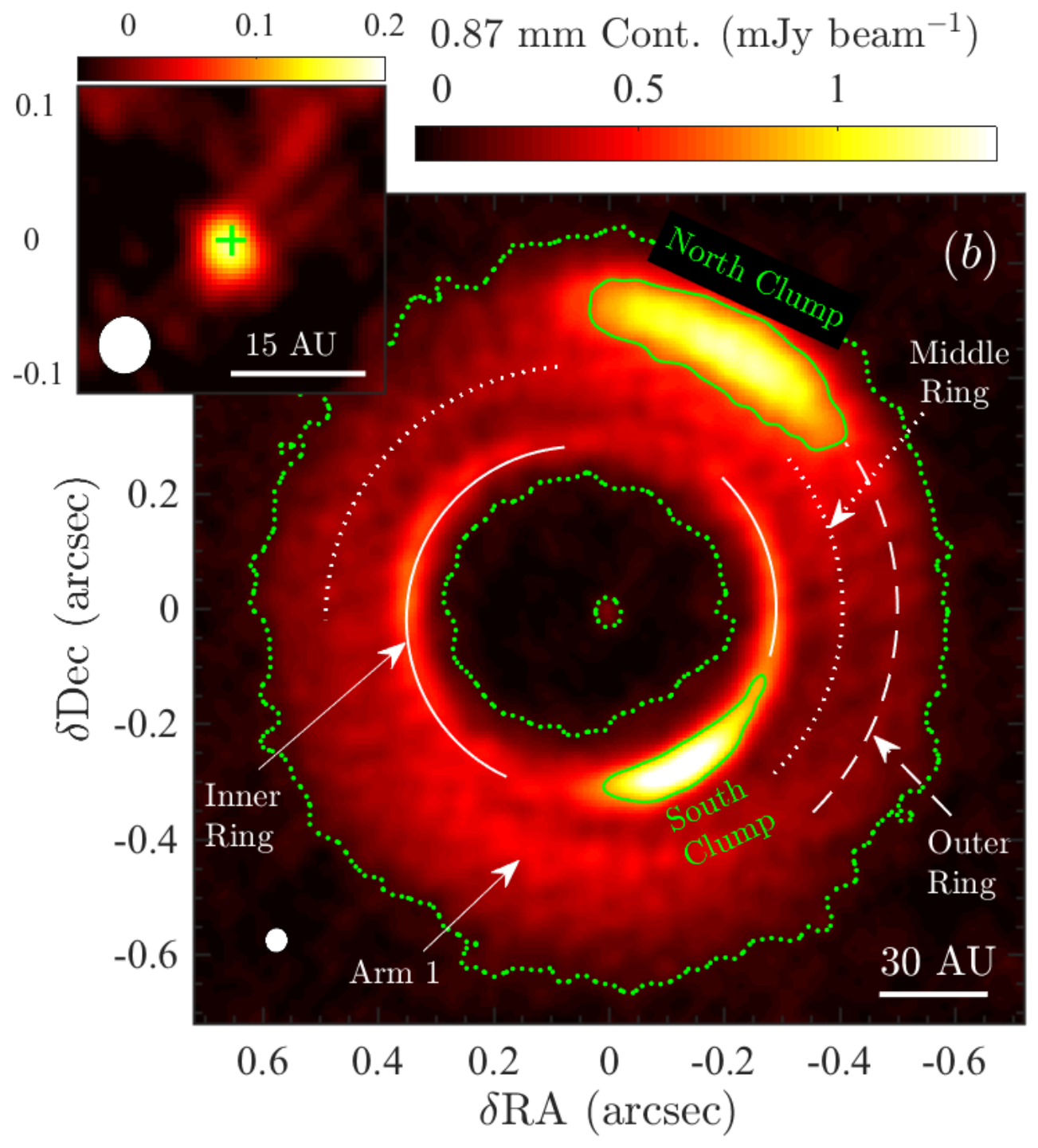}
\includegraphics[trim=0 0 0 0, clip,width=0.45\textwidth,angle=0]{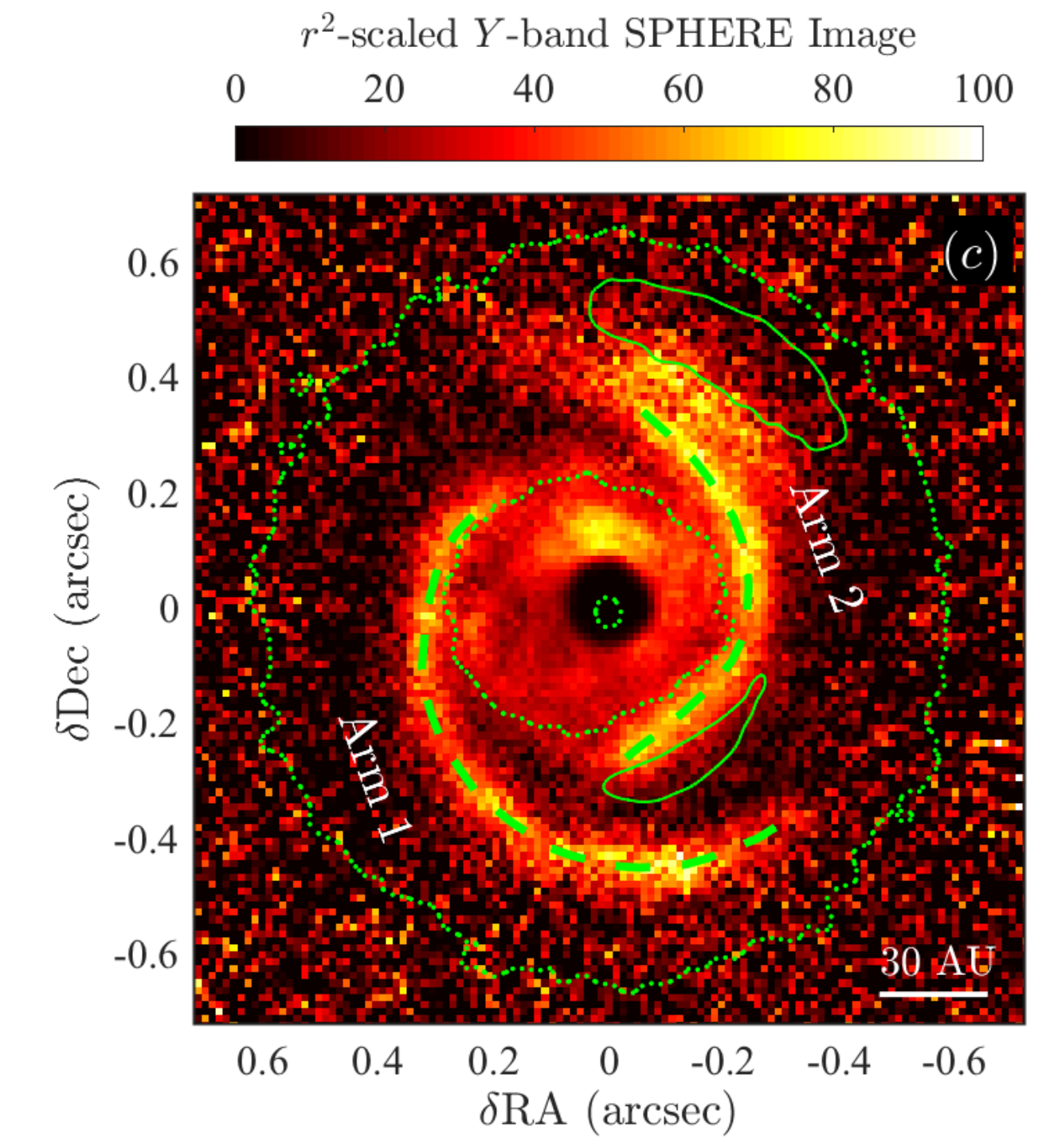}
\includegraphics[trim=0 0 0 0, clip,width=0.45\textwidth,angle=0]{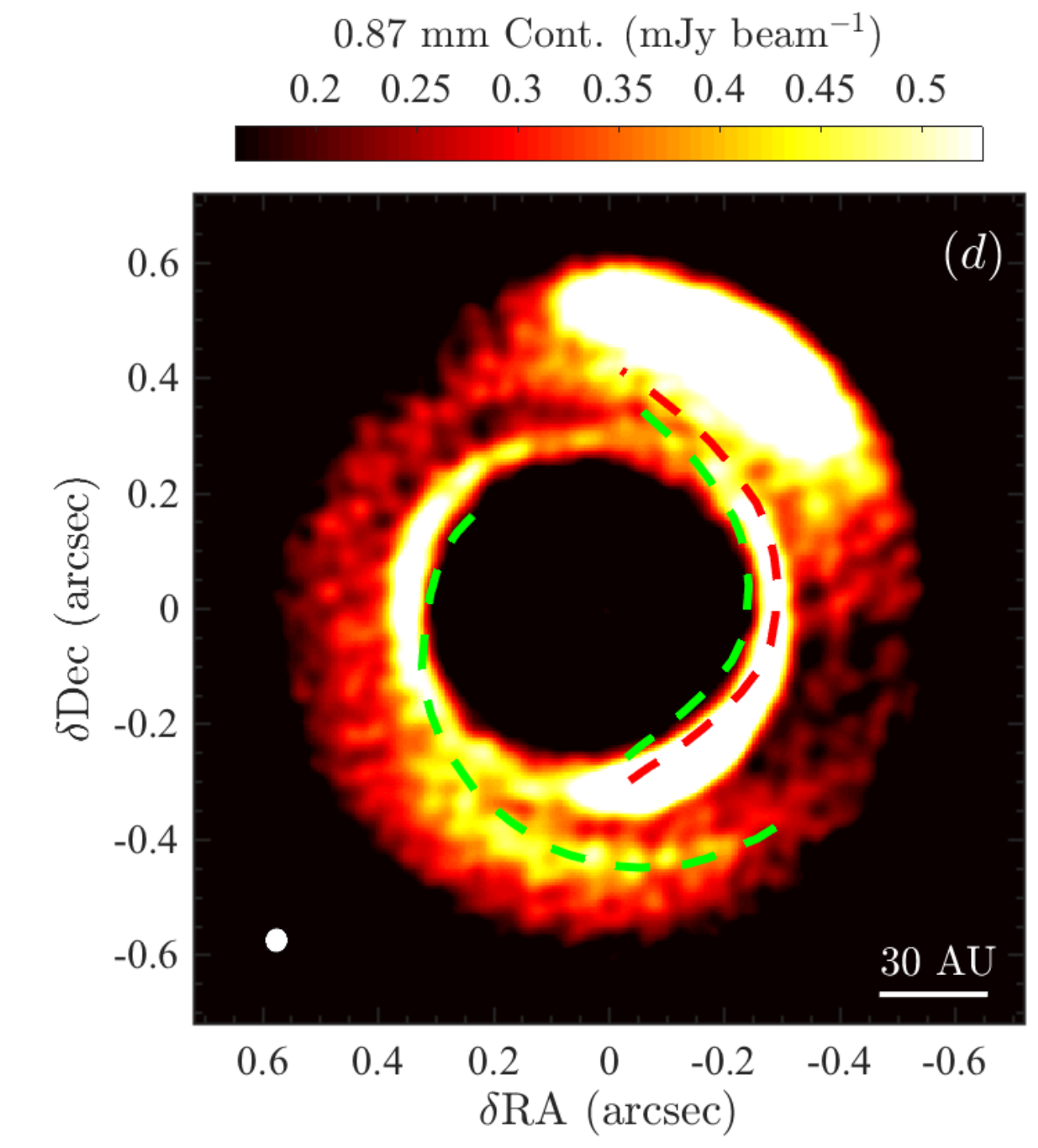}
\figcaption{\footnotesize Panels ($a$) and ($b$): ALMA 0.87~mm continuum emission from MWC 758 with a beam size 43$\times$39 mas (6.9$\times$6.2 AU; labeled at the lower left corner). North is up and east is to the left. In panel ($b$), structures are labeled; the green dotted contours are at the $3\sigma$ noise level; the two green solid contours are at half peak intensity at each clump; the solid, dotted, and dashed white arcs trace out the inner, middle, and outer rings, respectively; and the inset is a 0\farcs2 zoom of the central region (stellar location marked by the green plus). Panel ($c$) shows $r^2$-scaled SPHERE $Y$-band (1 $\micron$) polarized scattered light image \citep[normalized unit]{benisty15}. The contours in panel ($b$) are overlaid, and the two green dashed lines trace out the locus of the two spiral arms (labeled as Arm~1 and Arm~2; the two green curves are overlaid in ($d$) as well). The astrometrical alignment between the SPHERE and ALMA images is done by aligning the location of the central star. The absolute stellar position in the SPHERE image may be accurate to half the SPHERE pixel size (6 mas). Panel ($d$) shows the emission map with an aggressive color stretch to highlight the ALMA Arm~1. The red dashed curve is the locus of SPHERE Arm~2 shifted away from the star by 0\farcs05. The south part of SPHERE Arm~1 is revealed at a slightly larger stellocentric distance by ALMA. Arm~2 cannot be clearly identified in the continuum emission.
\label{fig:image}}
\end{center}
\end{figure}

\begin{figure}
\begin{center}
\text{Deprojected 0.87~mm Continuum Emission Map}\par\smallskip
\includegraphics[trim=0 0 0 0, clip,width=0.495\textwidth,angle=0]{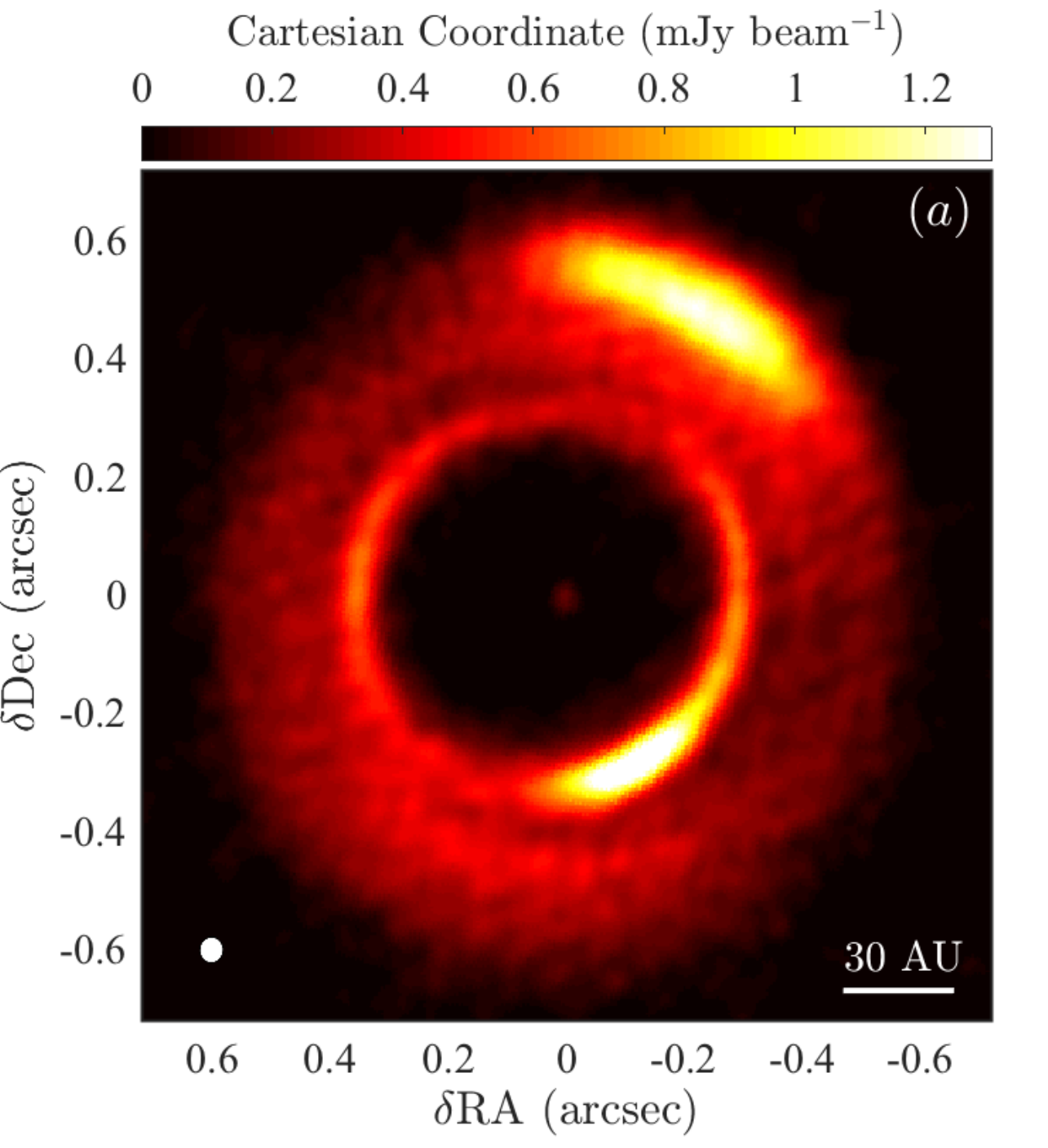}
\includegraphics[trim=0 0 0 0, clip,width=0.495\textwidth,angle=0]{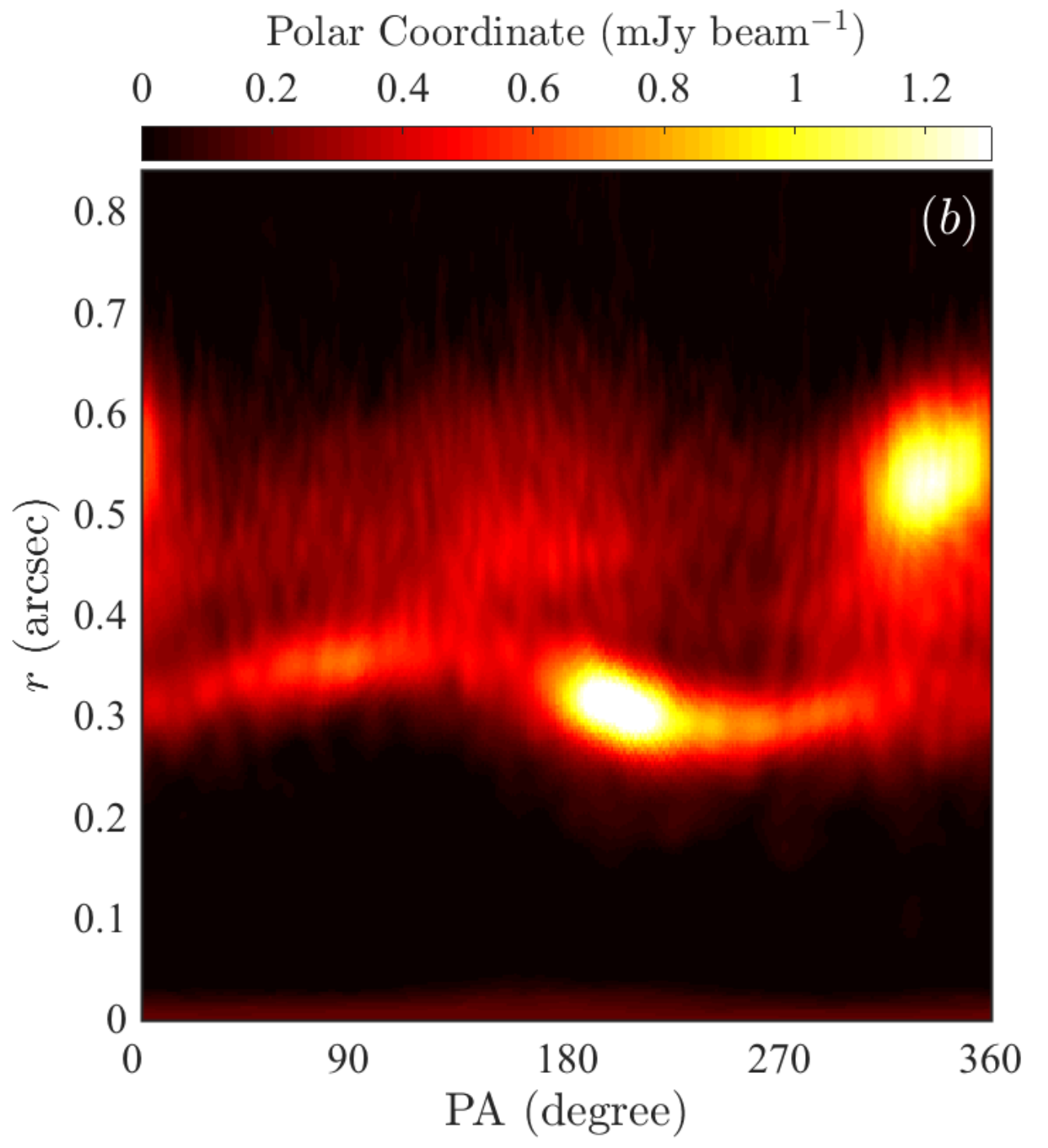}
\includegraphics[trim=0 0 0 0, clip,width=0.495\textwidth,angle=0]{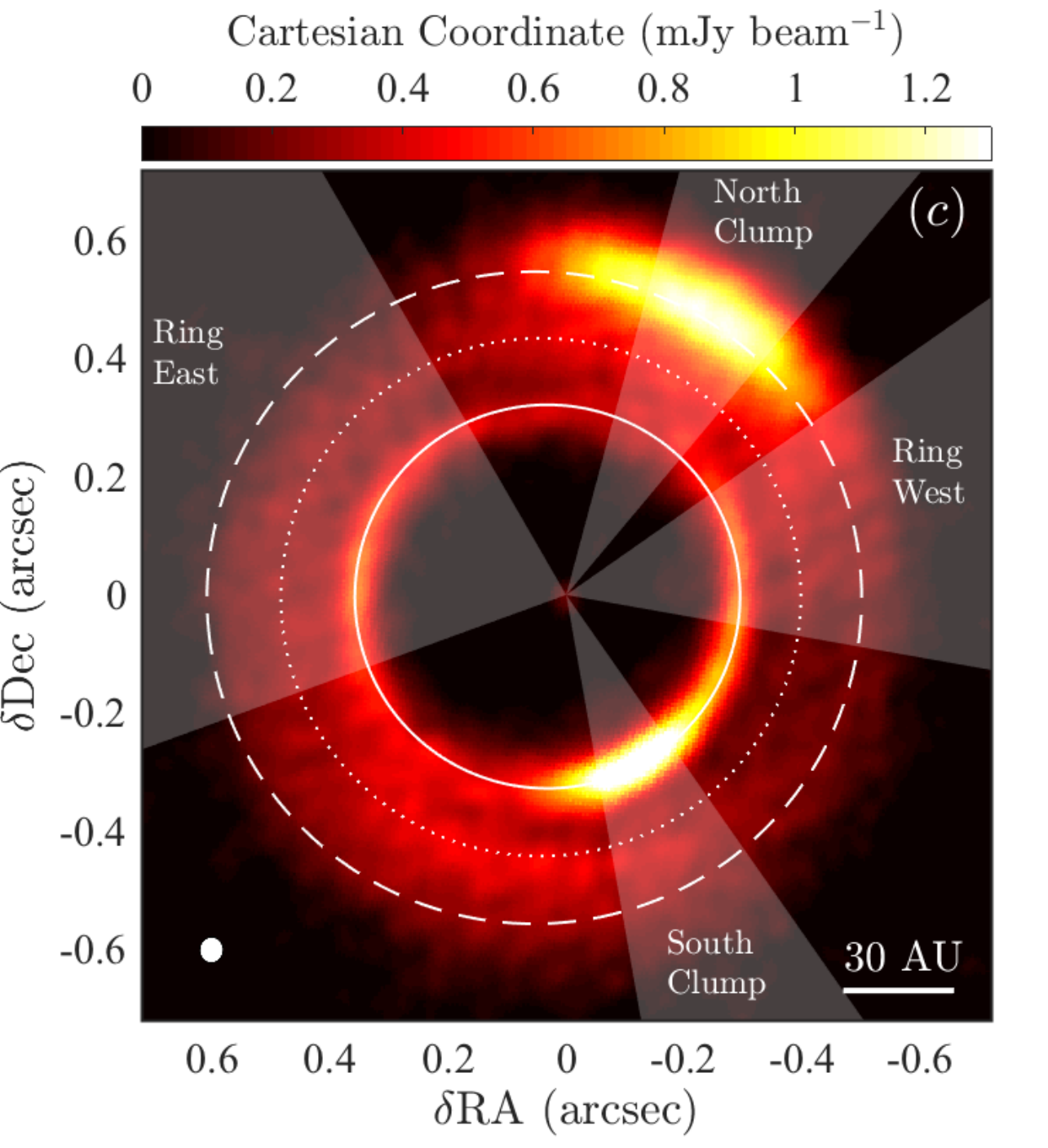}
\includegraphics[trim=0 0 0 0, clip,width=0.495\textwidth,angle=0]{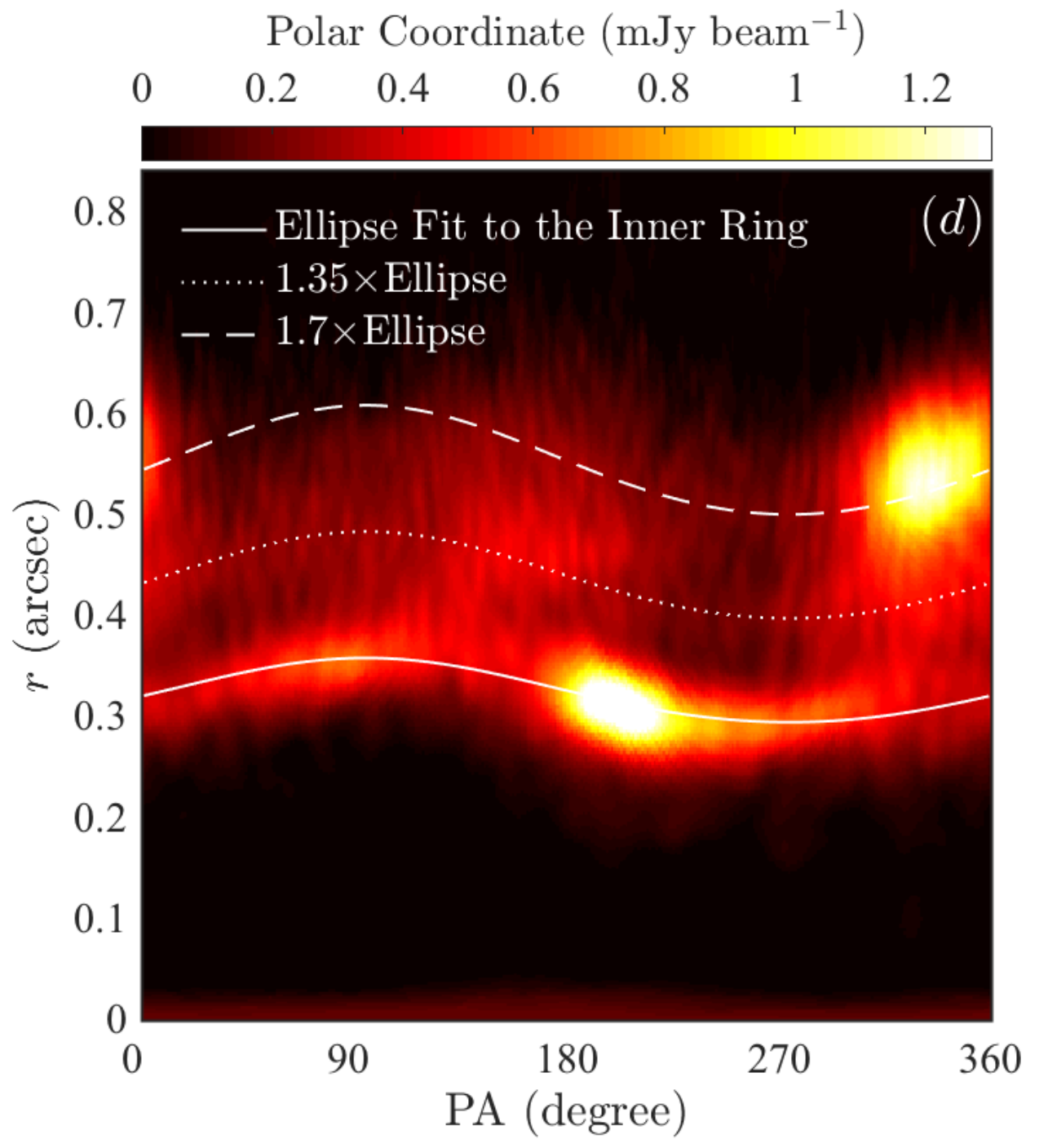}
\figcaption{\footnotesize Deprojected image of MWC 758 assuming $i$=21$^\circ$ and PA=62$^\circ$ in Cartesian and polar (radial-azimuthal) coordinates. The top and bottom rows are identical, apart from the annotations. In panel ($c$), the three curves are three ellipses 1$\times$, 1.35$\times$, and 1.7$\times$ the size of the ellipse fit to the inner ring at the cavity edge ($a$=0\farcs32, $e$=0.1, and one focus on the star), and the four shaded regions are selected azimuthal directions (their radial profiles are plotted in Figure~\ref{fig:deprojprofile}$a$ and \ref{fig:deprojprofile}$c$). In panel ($d$), the three curves are the three ellipses in panel ($c$) (line-types are consistent). They roughly trace out (inside to outside) the inner ring and the visible portion of the middle and outer rings marked in Figure~\ref{fig:image}$b$. The cavity is off-center and non-circular. See \S\ref{sec:results_cavity} and \S\ref{sec:cavity} for discussions. 
\label{fig:deproj}}
\end{center}
\end{figure}

The cavity and the narrow inner ring at the cavity edge in Figure~\ref{fig:image}$a$ appear to be non-circular and off-center. This is not a projection effect. Figure~\ref{fig:deproj} shows the deprojected view of the disk in Cartesian and polar (radial-azimuthal) coordinates assuming an inclination $i$=21$^\circ$ and position angle PA=62$^\circ$ \citep{isella10, boehler18}; the origin of the polar coordinate system is set to be at the expected GAIA stellar location. We note that the inclination and PA correspond to the outer disk outside the cavity. The system may have a mildly warped (misaligned) inner gas disk inside $\sim$0\farcs1, with an inclination in the range of 30--40$^\circ$ \citep{eisner04, isella08, boehler18}. Since $\sim$mm-sized dust traced by ALMA continuum emission is expected to settle to a thin layer at the disk midplane \citep[e.g.,][]{dullemond04dustsettling}, deprojection roughly recovers the face-on view of the disk. The non-circular and off-center cavity is clearly evident in deprojected maps --- circles centered on the star would be horizontal straight lines under the polar view. The deprojected inner ring is well-approximated by an ellipse with one focus on the star (note that the central point source roughly coincides with the GAIA stellar location, which is set to be the phase center), semi-major axis $a$=0\farcs319$\pm$0.002, eccentricity $e$=0.10$\pm0.01$, and major axis PA=95$^\circ$$\pm$10$^\circ$. 

Our conclusion that the cavity is not a circle is derived primarily from the inferred offset of the center of the ring from the stellar position. The deprojected inner ring may also be approximated by a circle with a radius $r$=0\farcs320$\pm$0\farcs003. However, its center is at ($\delta$RA, $\delta$Dec)=(33$\pm$2 mas, -5$\pm$3 mas) from the expected (GAIA) stellar position. This difference is $\sim$one order of magnitude larger than the expected ALMA astrometry uncertainty (\S\ref{sec:results_pointsources}, and \S A.9.5 in ALMA cycle 6 Proposal Guide). It is therefore very unlikely for the expected stellar location to offset from its actual location by $\sim$30 mas. In addition, the detection of the central sub-mm point source at roughly the expected stellar location suggests that it is within a few mas from the actual stellar location, as the point source is likely a small circumstellar disk. We also find no good physical motivation and consider it is very unlikely for the cavity to be a true circle but significantly off-centered. Future simultaneous high resolution observations of both the gas and the dust emission may provide more definitive evidence on the eccentric cavity.

Two depressions on the inner ring can be identified around PA$\sim$0$^\circ$ and 135$^\circ$, in between which are two bright arcs on the west and east sides. The ``Ring East'' and ``Ring West'' regions defined in Figure~\ref{fig:deproj}$c$ are relatively free from ``contaminations" from other features. We use them to study the radial structures (rings and gaps) in the outer disk at $r$$\sim$0\farcs25$-$0\farcs60. 

\begin{figure}
\begin{center}
\includegraphics[trim=0 0 0 0, clip,width=\textwidth,angle=0]{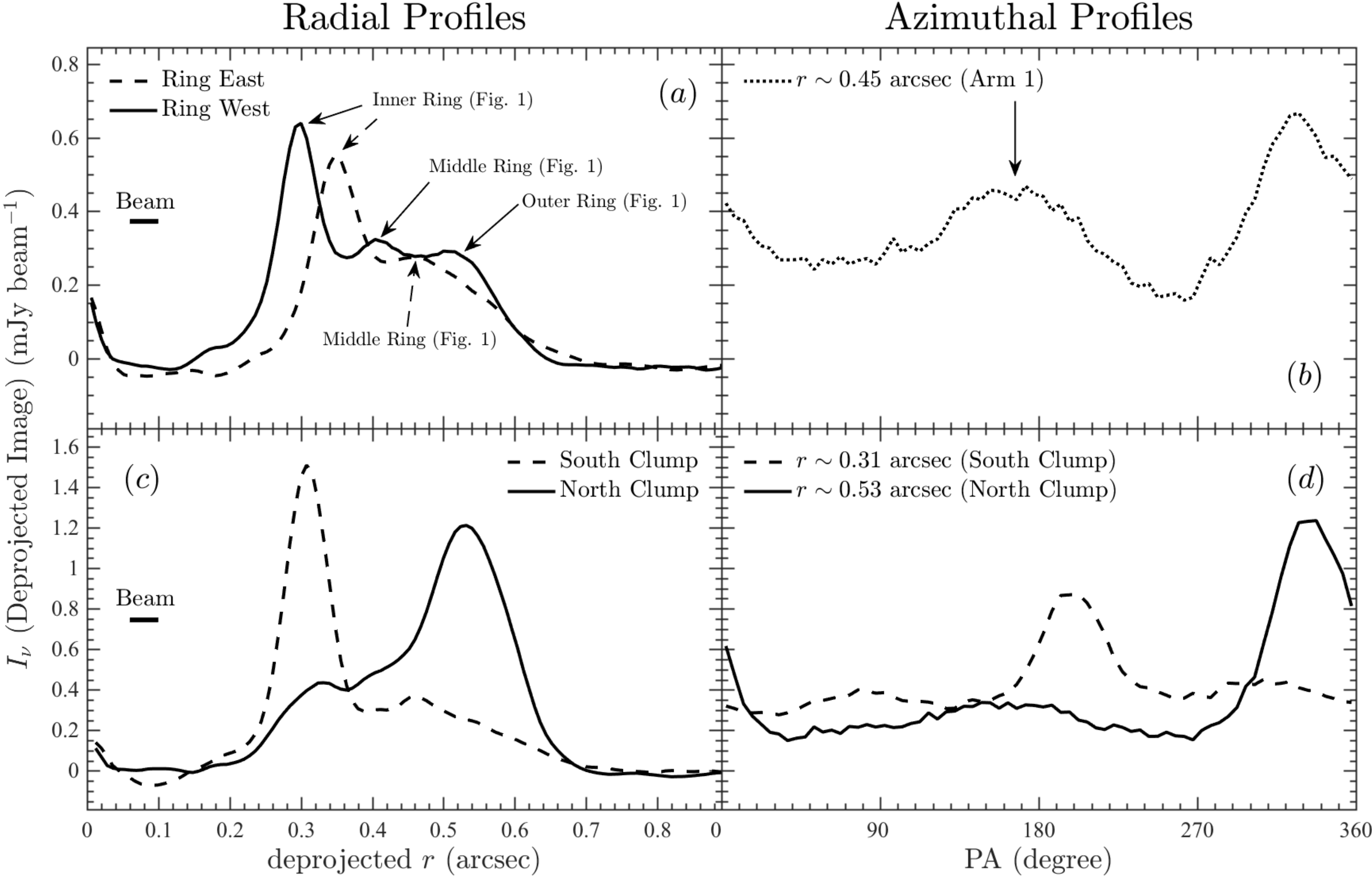}
\figcaption{Azimuthally-averaged radial profiles in four selected regions (($a$) \& ($c$)) and radially-averaged azimuthal profiles in three circular rings (($b$) \& ($d$); the radial widths of the rings are 0\farcs03 (Arm 1), 0\farcs04(north clump), and 0\farcs13 (south clump)) in the deprojected ALMA image (Figure~\ref{fig:deproj}). The four regions in ($a$) and ($c$) are defined as the shaded regions in Figure~\ref{fig:deproj}$c$, and the three rings in ($b$) and ($d$) are centered on the peak radii of Arm~1 and the two clumps. See \S\ref{sec:results_cavity} and \S\ref{sec:cavity} for discussions.
\label{fig:deprojprofile}}
\end{center}
\end{figure}

Figure~\ref{fig:deprojprofile}$a$ plots the azimuthally averaged radial profiles in these two regions. The Ring East region shows a global peak corresponding to the inner ring, and a second peak at 0\farcs45. The Ring West region shows the same inner ring at a smaller radius due to its eccentric nature, and two peaks in the outer disk at 0\farcs40 and 0\farcs51. We cannot trace out the 2$\pi$ extent of these outer rings as the emission is segmented into a west and an east part by the north clump and the region around Arm~1. Nevertheless, we tentatively associate the 0\farcs45 peak on the Ring East and the 0\farcs40 peak on the Ring West profile to a common faint middle ring (labeled in Figure~\ref{fig:deprojprofile}$a$ and Figure~\ref{fig:image}$b$). The dotted curve in Figure~\ref{fig:deproj}$c$ is an ellipse with the same orientation and eccentricity but 1.35$\times$ the size of the ellipse fit to the inner ring. It roughly traces out and connects the east (PA$\sim$220$^\circ-$310$^\circ$) and west (PA$\sim$10$^\circ$-120$^\circ$) segments of the middle ring (we suggest to view this panel together with the annotation-free version of the polar map, and also to view the ring in Figure~\ref{fig:image}d with a more dramatic color stretch). Similarly, the dashed curve is an ellipse 1.7$\times$ bigger; on the west side it roughly traces out the faint outer ring (labeled in Figure~\ref{fig:deprojprofile}$a$ and Figure~\ref{fig:image}$b$), while on the east side it passes through a region with S/N too low to enable robust structure identifications. These two rings merge into the north clump in the north end and touch the spiral region in the south. Both are unresolved in the radial direction, thus their physical widths are smaller than the beam size ($\sim$6.5 AU). The ``outer ring'' in the ``double-ring'' structure detected in a previous ALMA dataset \citep{boehler18} with 0\farcs1$-$0\farcs2 angular resolution may contain both the middle and outer rings in this dataset. All three rings (inner, middle, outer) may share similar eccentricities and orientations, although this needs to be confirmed by future observations.

\subsection{The Spirals}\label{sec:results_spirals}

MWC 758 has two prominent spiral arms in NIR scattered light (Arm~1 on the east and Arm~2 on the west, labeled in Figure~\ref{fig:image}$c$). Figure~\ref{fig:image}$d$ shows the map with an aggressive color stretch to highlight the spiral arms. Arm~1 is clearly detected at $r$$\sim$0\farcs4$-$0\farcs5 and PA$\sim$120$-$210$^\circ$. It is resolved in the azimuthal but not the radial direction, setting an upper limit on the physical radial-width-to-radius ratio to be 10\%. The east end of the arm starts at the inner ring. The arm weakens and merges into the background at the west end. The peak surface brightness on the arm is 0.51 mJy beam$^{-1}$ (S/N=25), reached at $\sim$72 AU (deprojected; 0\farcs45), corresponding to a brightness temperature $T_{\rm B}$=9.0 K. The azimuthal peak-to-trough contrast of the spiral in the deprojected image measured along the $r$=0\farcs45 circle (no radial or azimuthal averaging) is 3.6. Figure~\ref{fig:deprojprofile}$b$ shows the azimuthal profile of the deprojected surface brightness averaged over a ring with radius of 0\farcs45 and width of 0\farcs03. Arm~2 is not clearly detected. However, it is possible that part of that arm is present and spatially overlapping with other features (see \S\ref{sec:arm2}).

\subsection{The Double Emission Clumps}\label{sec:results_clumps}

The two emission clumps are resolved in both the azimuthal and the radial directions (see the beam size and the half-peak contours around the clumps in Figure~\ref{fig:image}$b$). Figure~\ref{fig:deprojprofile}$c$ shows the radial profiles across the two clumps. The radial full width at half maximum (FWHM) of the south and north clumps in the deprojected image are $\sim$1.8 and $
\sim$3.9 times the beam size, respectively, and the azimuthal FWHM of both clumps is $\sim$1/6 of a circle. We estimate their intrinsic (deconvolved) radial FWHM (FWHM$_{\rm intrinsic}$$\approx$$\sqrt{{\rm FWHM}^2-{\rm FWHM_{\rm beam}^2}}$) to be 0\farcs063 and 0\farcs155 for the south and north clumps, respectively. Figure~\ref{fig:deprojprofile}$d$ shows the azimuthal profiles of the clumps in the deprojected map. The azimuthal peak-to-trough contrasts of the south and north clumps are 4.4 and 10 at the radii of their peaks, respectively. The peak surface brightnesses in the south and north clumps are 1.60 and 1.28 mJy beam$^{-1}$, corresponding to $T_{\rm B}$=16.8 and 14.7 K, respectively. 

We generate representative disk models using the radiative transfer code HOCHUNK3D \citep{whitney13} to obtain crude estimates of the dust temperature $T_{\rm dust}$ at these radii. The models largely follow the radiative transfer model for MWC 758 presented in \citet{grady13}. They assume a full disk with surface density $\Sigma\propto1/r$, as the $\micron$-sized dust probed by scattered light is present inside the cavity \citep[the starlight is mainly absorbed and scattered by the small dust]{benisty15}. The central star is assumed to have $T_\star$=7580 K and $L_\star$=9.7$L_\odot$ after scaling the stellar luminosity in \citet{andrews11} by the new GAIA distance. With a variety of disk scale heights ($h$=5 AU to 20 AU at $r=$100 AU), and disk flaring angles ($h/r\propto r^{0.05}$ to $h/r\propto r^{0.25}$), we obtain $T_{\rm dust}$=20$-$32 K at the location of the south clump (deprojected $r$=0\farcs31=50 AU), and 17$-$27 K at the location of the north clump (deprojected $r$=0\farcs53=85 AU). The two clumps may be marginally optically thin (vertical optical depth of order unity). The same models also yield $T_{\rm dust}$=18$-$27 at the location of Arm~1 ($r$$\sim$70 AU after deprojection), significantly higher than the $T_{\rm B}$ (9.0 K).


\section{Discussion}\label{sec:discussions}

\subsection{The Eccentric Cavity and Rings}\label{sec:cavity}

Rings with eccentricities $e \gtrsim 0.1$ have been found or 
inferred in debris disks \citep{kalas05,lee16chiang}. To our
knowledge, MWC 758 as revealed in our ALMA dataset provides
the first known example of an intrinsically eccentric
protoplanetary disk.

One hypothesis for the origin of transitional disks is that their
cavities are opened by (multiple) companions, possibly planets. 
Simulations have shown that the outer edge of a gap opened by a 
companion on a circular orbit may become eccentric 
\citep[e.g.,][]{kley06, dunhill13, farris14,teyssandier17}.
The gap edge eccentricity is excited by interactions between 
the companion and the disk at the outer 1:3 Lindblad resonance 
(LR), and damped by the 1:2 LR and the co-orbital resonance; 
therefore, a gap wide enough to extend past the outer 1:2 but
not the 1:3 LR may develop an nonzero eccentricity 
\citep{kley06}. A higher companion mass, a lower disk viscosity, 
and a lower disk temperature facilitate the opening of deep and 
wide gaps \citep{fung14}, and thus the growth of $e$. 
In this scenario, the outermost planet inside the cavity should 
be located between 0.48--0.63 of the cavity outer edge in the 
gas, or 25--32 AU assuming that $r_{\rm gas,\ cavity} = r_{\rm dust,\ cavity} = 0\farcs32 = 51$ AU.
Since the cavity size in the gas is expected to be smaller than it is in dust emission due to gas-dust coupling at the gas cavity edge \citep{zhu12, pinilla12diffcavsize}, the planet may be at smaller stellocentric radius.

The eccentric cavity and the outer disk precess at a frequency much lower than the local Keplerian frequency $\Omega_{\rm K}$ \citep[e.g.,][]{teyssandier17}. \citet{hsieh12} showed that in this case the eccentric outer disk does not trap $\sim$mm-sized particles in the azimuthal direction. Thus we do not expect the two observed emission clumps to be global dust traps produced by the eccentric disk. Instead, they may be generated by other mechanisms, such as dust trapping by vortices (see \S\ref{sec:clump}). An eccentric gap may coexist with vortices in disk-planet interaction models \citep{ataiee13}.

Eccentric cavities and rings can alternatively be opened by companions on eccentric orbits. We note that giant planets with masses of several $\mj$ interacting with a gaseous disk may grow their eccentricities to $\sim$0.1, comparable to the eccentricity observed here \citep{dunhill13, duffell15eccentricplanet, ragusa17}. In this scenario, the orbits of the cavity-opening planet(s) and gap-opening planet(s) in MWC 758 may share similar eccentricities and orientations. Future studies are needed to investigate this hypothesis.

\subsection{The Spiral Arms}\label{sec:spiral}

Among the new features, perhaps the most exciting one is the discovery of the sub-mm continuum counterpart to the NIR spiral on the south. Spiral arms in protoplanetary disks have been detected in NIR scattered light \citep[e.g.,][]{fukagawa06, hashimoto11, muto12, garufi13, canovas13, currie14, wagner15hd100453, akiyama16, stolker16sao206462, liu16fuori, maire17, benisty17, avenhaus17, long17, follette17, canovas18, langlois18}, and also in $\sim$mm gas \citep[e.g.,][]{corder05, christiaens14, tang17} and dust emission \citep[e.g.,][]{perez16, tobin16}. A few mechanisms have been explored to explain their origin, including companion-disk interaction \citep[e.g.,][]{dong15spiralarm, zhu15densitywaves, bae16sao206462}; gravitational instability \citep[GI; e.g.,][a combination of GI + planet, \citealt{pohl15}]{dong15giarm, hall16, dipierro15gidisks, dong16protostellar, tomida17, meru17, hall18}, finite light travel time \citep[e.g.,][]{kama16}, and moving shadows \citep[e.g.,][]{montesinos16, montesinos17}. For planet-induced spiral arms, their morphology and brightness can be used to constrain the mass of the perturber \citep{fung15, dong17spiralarm}. 

Except in rare cases (e.g., HD 100453, in which the spiral arms are driven by a visible stellar mass companion bound to the system; \citealt{dong16hd100453, wagner18}), the true origin of observed spiral arms, in particular whether they are planet-induced, is under debate. For MWC 758, the low disk mass ($M_{\rm disk}$) estimated from sub-mm continuum emission \citep[$M_{\rm disk}$$\sim$1\%$M_\star$;][]{andrews11} disfavors the GI scenario, which usually requires $M_{\rm disk}$$\gtrsim$10\%$M_\star$ under typical conditions \citep{kratter16}. Spatially resolved observations of spirals revealing their shapes, contrasts, and locations at multi-wavelengths are crucial to a thorough understanding of their origins. 

Companion-induced spiral arms are pressure waves. They only directly manifest in a pressure supported medium --- the gas. Whether they are present in sub-mm continuum emission depends on whether they can trap $\sim$mm-sized dust particles. Such spirals, co-rotating with their drivers, have a nonzero relative angular velocity with respect to the local disk, different from GI-induced spirals. While the latter are capable of trapping particles of certain sizes \citep[e.g.,][]{rice04, dipierro15gidisks}, whether and how companion-induced spiral arms can trap dust are uncertain. Further, while the primary arm (the one directly pointing to the companion) has been thoroughly studied \citep[e.g.,][]{goldreich79, ogilvie02, goodman01, rafikov02}, the excitation, properties, and propagation of the secondary (and additional) arm driven by a single companion is still ongoing research \citep[e.g.,][]{juhasz15, lee16, hord17, juhasz17, bae18theory, bae18simulation, arzamasskiy18}. 

With the nature of the two arms still being uncertain, the large azimuthal peak-to-trough contrast of Arm~1 in the deprojected sub-mm continuum emission suggests it is mainly a density (emission optical depth) feature. The spiral is unlikely to be a shadow or temperature feature, as no evidence of the shadow is seen in NIR scattered light. If dust is optically thin and the structure is spatially resolved, the (sub-)mm emission intensity $I_\nu$ at a frequency $\nu$ is proportional to $\Sigma_{\rm dust}$$\times$$\kappa_{\nu,\rm dust}$$\times$$B_\nu(T_{\rm dust})$, where $\Sigma_{\rm dust}$ and $\kappa_{\nu, \rm dust}$ are the surface density and opacity of the sub-mm-continuum-emitting dust, and $B_\nu(T_{\rm dust})$ is the Planck function at the dust temperature $T_{\rm dust}$. Note that Arm 1 is not resolved radially in our ALMA observations, and the measured azimuthal contrast is only the lower limit of the ``intrinsic'' azimuthal contrast in $I_\nu$ with infinite angular resolution. If Arm 1 is caused by variations in $T_{\rm dust}$ as in the shadow scenario \citep{montesinos16, montesinos17}, at least order unity variation in $T_{\rm dust}$ in the azimuthal direction across the arm is needed, which seems physically unlikely. Instead, the arm is most likely introduced by variations mainly in $\Sigma_{\rm dust}$$\times$$\kappa_{\nu,\rm dust}$, with a possible minor contribution from $T_{\rm dust}$ \citep[e.g., from spiral shock heating;][]{rafikov16}.

\subsubsection{Why ALMA Arm~1 is Offset from NIR Arm~1}\label{sec:armlocation}

\begin{figure}
\begin{center}
\includegraphics[trim=0 0 0 0, clip,width=\textwidth,angle=0]{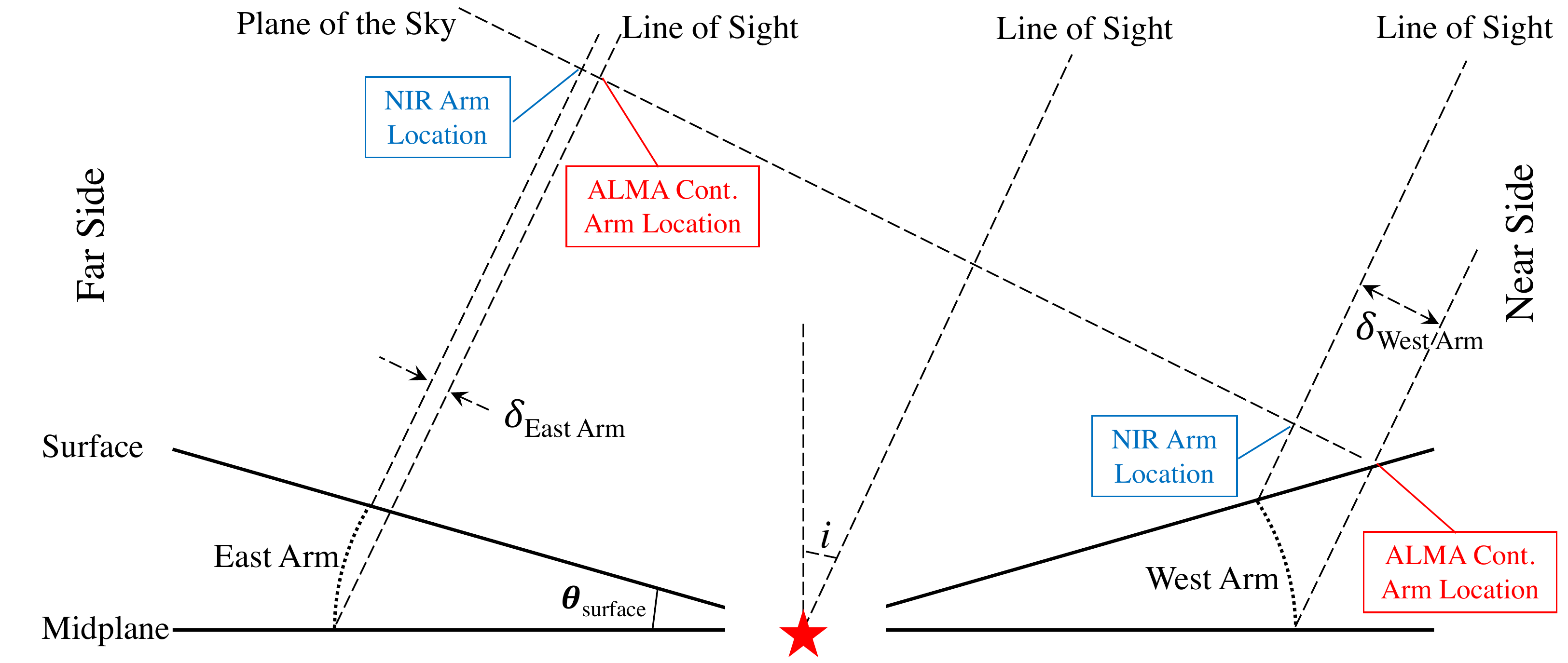}
\figcaption{A schematic of ALMA and NIR imaging observations of a disk with two planet-induced spiral arms viewed from the side. The observer is to the northwest. The same spiral arm can be projected to different locations on the plane of the sky in NIR scattered light (probing surface features) and $\sim$mm continuum emission (probing midplane features). The East Arm in NIR scattered light is at a larger distance from the star than it is in ALMA continuum emission, while it is the opposite for the West Arm. This is caused by both the curved vertical structure of spiral arms (illustrated by the dotted curves) and the inclination of the disk. See \S\ref{sec:spiral} for discussions.
\label{fig:schematic}}
\end{center}
\end{figure}

At PA$\sim$120$^\circ-$180$^\circ$, Arm~1 in the ALMA continuum emission map is at a slightly larger stellocentric distance than in NIR scattered light (Figure~\ref{fig:image}$d$; maximum radial offset $\sim$0\farcs03, or 8\% of the radius). The ALMA arm crosses the NIR arm at PA$\sim$180$^\circ$, and the two roughly overlap at PA$\sim$180$^\circ-$210$^\circ$. This is most likely caused by the difference between a midplane feature and a surface feature in a projected view, as illustrated in the schematic in Figure~\ref{fig:schematic} (see also \citealt{stolker16}). When viewed at a finite inclination $i$, midplane structures probed by ALMA continuum observations are simply compressed by $1/\cos{i}$ along the direction of the minor axis. Such images are  ``deprojectable''. On the other hand, scattered light come from a cone-shaped disk surface. \citet{dong16armviewing} showed that surface features at a finite inclination can be dramatically distorted from their morphology at face-on, and such distortions cannot be recovered by simple deprojection. When viewed at nonzero $i$, ``NIR / surface spiral" and ``ALMA / midplane spiral" are projected to different locations on the plane of the sky. An additional complication is that spiral arms are curved in the vertical direction --- they bend over towards the star (Fig. 1, \citealt{zhu15densitywaves}; Fig. 2, \citealt{lyra16}; dotted curves in the schematic). In the schematic, the surface (NIR) East Arm is further from the star than the midplane (ALMA) East Arm on the plane of the sky, while for the West Arm it is the opposite. 

\begin{figure}
\begin{center}
\includegraphics[trim=0 0 0 0, clip,width=0.495\textwidth,angle=0]{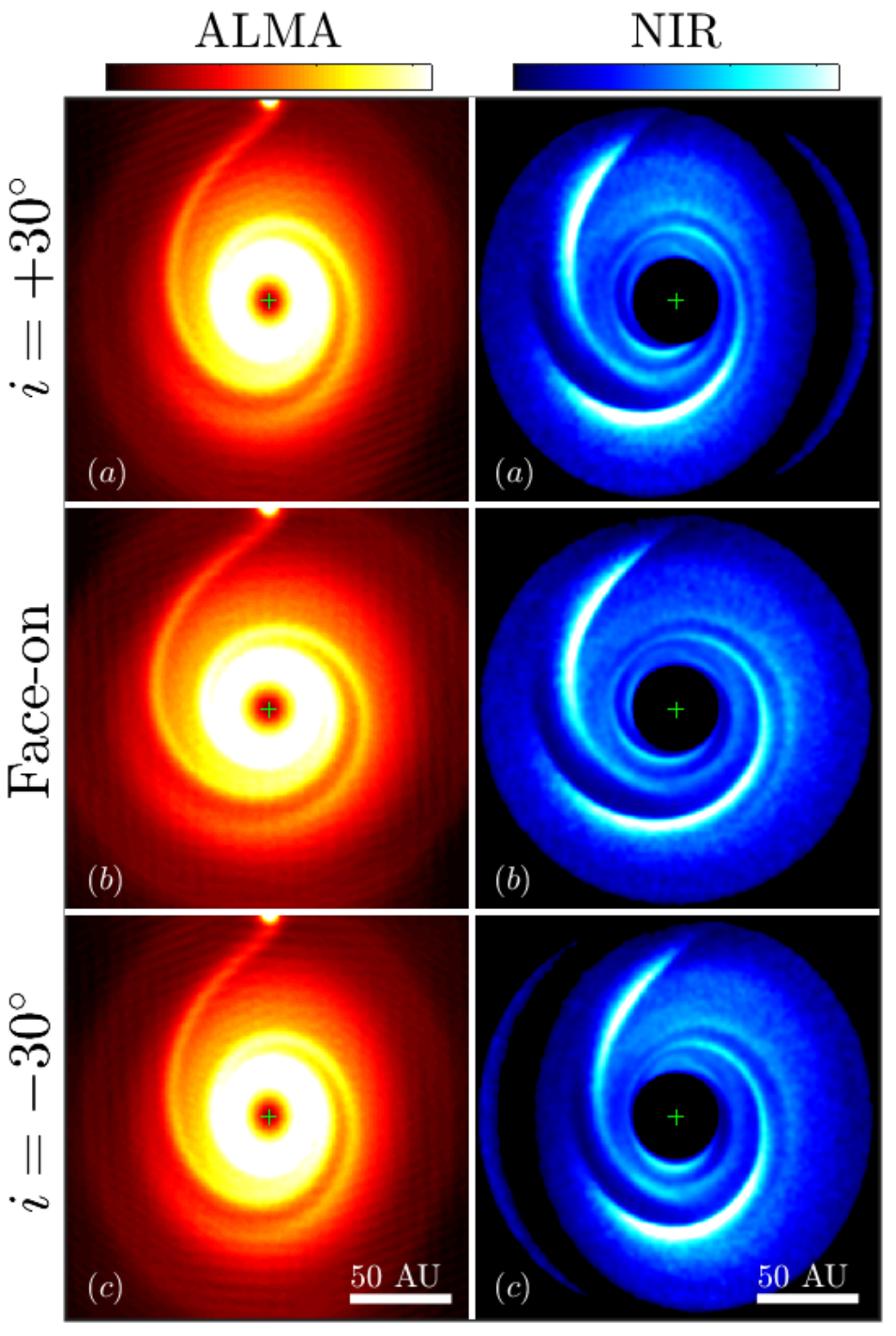}
\includegraphics[trim=0 0 0 0, clip,width=0.495\textwidth,angle=0]{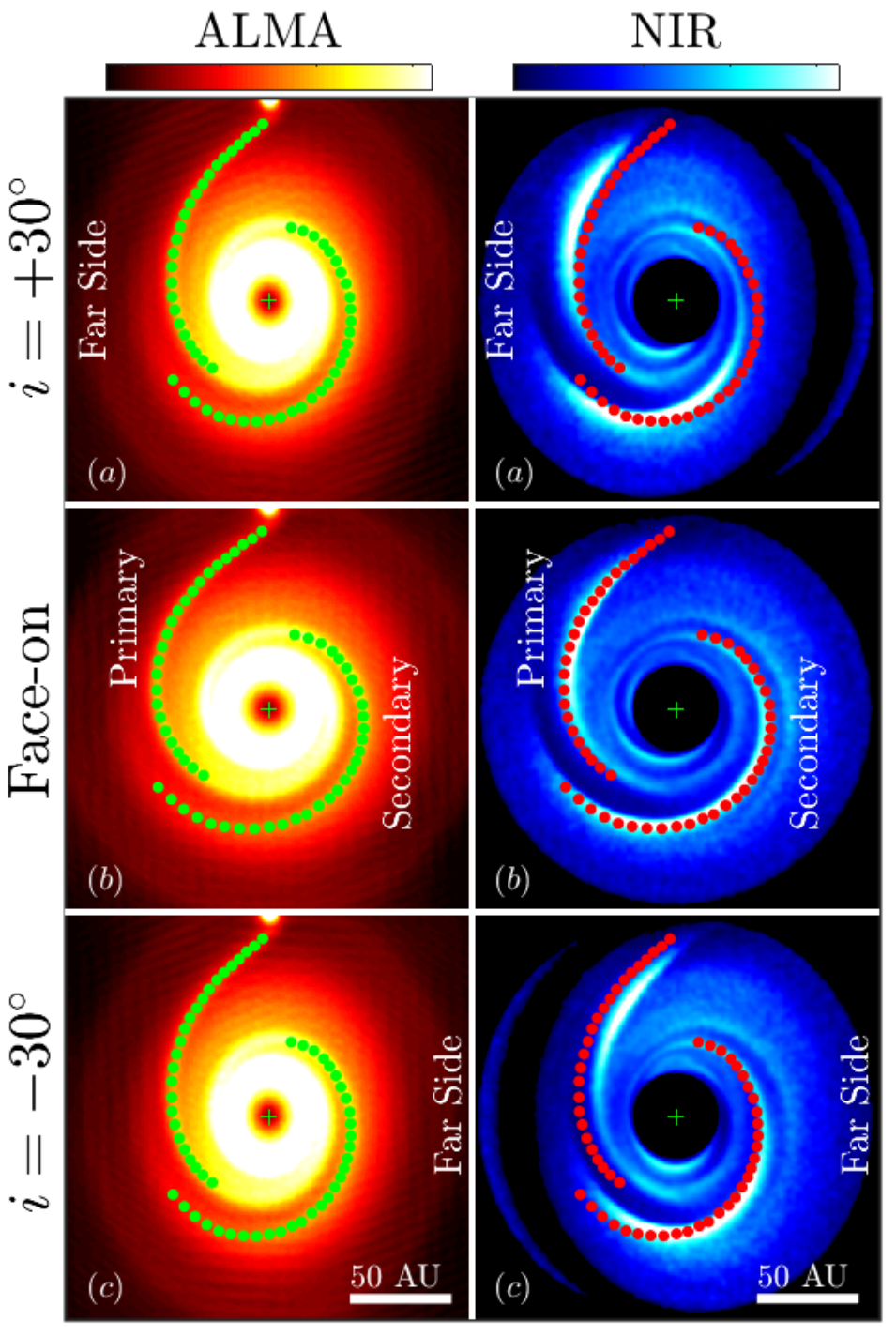}
\figcaption{Synthetic ALMA 0.87 mm continuum emission and NIR ($H$-band) polarized scattered light images (central part artificially masked) of a toy disk model with two spiral arms excited by a planet ($\mplanet=3\times10^{-3}M_\star$; located at $r$=100 AU and PA=0) at three viewing angles (inclination $i$ labeled on the left; PA=0 in all cases, i.e., the major axis is along North-South), produced using the $3\mj$ model in \citet{dong16armviewing}. The disk is assumed to be at 151 pc and images at both wavelengths are convolved to have an angular resolution of 0\farcs04. Color stretch is linear and units are arbitrary. $H$-band images have been scaled by $r^2$ and ALMA images have been scaled by $\sqrt{r}$ (deprojected $r$) to highlight the arms. The right half is identical to the left half, but with (1) the locus of the two arms in the ALMA image at each viewing angle marked by the dots in both the ALMA and NIR images (the green and red dots are the same), (2) the far side of the disk marked in the top and bottom rows, and (3) the primary and secondary arms labeled in panels ($b$). The sub-mm continuum emitting big dust ($\sim$mm-sized) are assumed to have surface density linearly scaled with the gas, and have settled to the disk midplane. The locations of arms in ALMA continuum observations (midplane features) may differ from where they are in NIR observations (surface features). See \S\ref{sec:spiral} for discussions.
\label{fig:simulation}}
\end{center}
\end{figure}

We present a toy model in Figure~\ref{fig:simulation} to visualize the effect. Synthetic ALMA sub-mm continuum and NIR polarized scattered light images for one disk model are simulated at three viewing angles using 3D hydrodynamics and radiative transfer simulations. The disk model is taken from the Model $3\mj$ in \citet{dong16armviewing}, which has a planet with $\mplanet=3\times10^{-3}M_\star$ on a circular orbit at $r=100$ AU, driving two main spiral arms interior to its orbit. The outer disk at $r>100$ AU is removed in post-processing to highlight the spiral arms. In simulating the ALMA images, continuum emitting dust is assumed to be mm in size. While mm-sized dust is not included in the gas-only hydro simulation, their surface density is assumed to be linearly proportional to the gas. The total mm-sized dust mass is normalized such that the disk is optically thin at sub-mm wavelengths. In the vertical direction the density distribution of mm-sized dust is assumed to be Gaussian, with the aspect ratio $h/r$, taken to be 10\% of the gas, less than 1\% everywhere. We emphasize that proper dust-gas coupling (i.e., realistic trapping of $\sim$mm-sized dust by the spiral arms) is not taken into account, and the toy model is only meant to visualize the ALMA-NIR (equivalently midplane-surface) arm location offset effect.

At $i=0$ (panels ($b$)), both the primary and secondary arms are located roughly at the same locations in ALMA and NIR images\footnote{In principle, even at face-on we may still expect a small difference as the arms tend to bend over towards the star in the vertical direction (see the schematic). In practice, this difference may be small. Usually the disk aspect ratio $h/r$ is smaller than 0.1 at tens of AU. Assuming the disk surface is at three scale heights, the opening angle of the surface from the midplane $\theta_{\rm surface}\lesssim\arcsin{(3\times0.1)}=17^\circ$, and $\cos{17^\circ}=0.96$. If in the vertical direction the arms bend to maintain a constant distance to the star, we expect $\lesssim$4\% midplane-surface arm location difference at $i=0$, or $\sim$0\farcs01 at 60 AU at 140 pc.}. When inclined, on the nearside of the disk, the surface feature moves towards the star faster than the corresponding midplane feature, thus NIR arms are found at shorter stellocentric distances than ALMA arms (the primary in ($c$), the east side of the secondary in ($c$), and the west side of the secondary in ($a$)). On the far side, NIR arms stay roughly at the same locations while ALMA arms approach the star, thus displacing the two (the primary in ($a$), the east side of the secondary in ($a$), and the west side of secondary in ($c$)). Also note that in NIR images a weak third arm right outside the secondary is present. This arm is mainly a surface feature caused by varying scale height, not a surface density feature, therefore not visible in ALMA images.

The difference in the arm locations in NIR and ALMA images ($\delta$ in the schematic) depends on (1) $i$, (2) PA, (3) the opening angle of the scattered light surface $\theta_{\rm surface}$, and (4) the arm structure in the vertical direction. Once (4) is thoroughly understood, an accurate measurement of $\delta$ combining with a known disk inclination may constrain $\theta_{\rm surface}$. If multi-wavelength scattered light images (e.g., from optical to $L$-band) are available, arms may move slightly across wavelengths as $\theta_{\rm surface}$ drops with increasing wavelength. Finally, we note that \citet{boehler18} detected Arm~1 in ALMA $^{13}$CO emission with a coarse beam ($\sim$0\farcs2), and concluded that it coincides with the NIR Arm~1. Future gas observations with higher angular resolution are needed to more accurately determine the location of the arm in the gas. Tentatively we regard the \citet{boehler18} result  as evidence for $^{13}$CO emission originating from a layer close to the disk surface. 

\subsubsection{Where is Arm~2 in Sub-mm Emission}\label{sec:arm2}

Arm~2 is not clearly revealed in our ALMA data. This may be due to less effective trapping of $\sim$mm-sized dust by Arm~2 compared to Arm~1. Specifically, if the two arms are produced by one planet on the outside \citep[the hypothesis raised by][see the discussion in \S\ref{sec:planet}]{dong15spiralarm}, they are expected to have different amplitudes in gas surface density (Fig. 5, \citealt{zhu15densitywaves}; Fig. 1, \citealt{fung15}; Fig. 2, \citealt{dong16armviewing}; see also \citealt{juhasz15}). This effect probably leads to different dust trapping capacities, resulting in different arm contrasts, locations, and morphologies in sub-mm continuum emission between the two arms. 

Another possibility is that Arm~2 is present but coincides with other features. As shown in Figure~\ref{fig:image}$d$, the south clump and the west part of the inner ring at PA$\sim$220$^\circ$$-$320$^\circ$ are located just outside the NIR Arm 2, and the north clump contributes emission at the location of the NIR Arm 2 at PA$\sim$320$^\circ-$360$^\circ$. Perhaps some of the emission along a track $\sim$0\farcs05 outside the NIR Arm~2 (the red dashed line) comes from an underlying arm structure that cannot be separately identified. Future disk modeling that decomposes the emission in this region to the relevant components are needed to elucidate the situation.

If (1) the MWC 758 disk's two arms are different in sub-mm continuum emission (e.g., one is present while the other is not) despite their relative symmetry in NIR scattered light, and (2) planet-induced primary and secondary arms do trap dust differently, then a careful comparison between ALMA observations and gas+dust simulations of disk-planet interactions may reveal which arm is the primary arm pointing to the hypothetical arm-driving planet. Previous simulations have shown that for multi-$\mj$ planets the two arms driven by a single planet are nearly equally prominent in scattered light \citep{dong15spiralarm, dong16armviewing}, making it difficult to break the degeneracy (unless the two arms are found in a near face-on disk and are not symmetric; \citealt{fung15}).

Similar to the second interpretation above, the east side of the inner ring at PA$\sim$30$^\circ-$120$^\circ$, being slightly outside NIR Arm~1, may also contain contributions from a north segment of ALMA Arm~1 as the two coincide. In this scenario, the ``intrinsic'' Arm~1 in sub-mm continuum emission would start all the way north at PA$\sim$30$^\circ$, covering a full $\sim$180$^\circ$ in the azimuthal direction similar to the NIR Arm~1. The azimuthal asymmetry of the inner ring might be partially induced by the ``extra'' contributions from the underlying spiral arms. 

\subsection{Dust Distributions in the Emission Clumps}\label{sec:clump}

As both clumps are spatially resolved and marginally optically thin, dust particles contributing to 0.87 mm continuum emission at each clump are distributed in an extended region, with their surface density roughly traced by the surface brightness of the emission. Hydro simulations of dust trapping by $m$=1 azimuthally asymmetric structures in the gas \citep[e.g.,][]{mittal15, baruteau16, miranda17} generally show effective trapping, resulting in compact dust concentrations if given enough evolution time. Particles of sizes in a finite range, all contributing to sub-mm continuum emission at a single wavelength, can be collected at different azimuthal locations inside one clump, resulting in extended emission over a large azimuthal angle, consistent with our observations \citep[e.g.,][]{mittal15}. In the radial direction, however, particles tend to drift into arcs narrower than the disk scale height, which is usually around 10\% of the radius or less. The radial drift timescale depends on the particle size. It is shortest --- shorter than system ages --- for particles whose dimensionless stopping times (Stokes numbers St) are on the order of unity.

At the two clumps, the average midplane temperature from the group of representative radiative transfer models introduced in \S\ref{sec:results_clumps} (26K and 22K for the south and north clumps, respectively) correspond to a local disk aspect ratio $h/r$=0.059 (south) and 0.071 (north) (assuming a 1.4$M_\odot$ stellar mass; \citealt{boehler18}). The intrinsic radial FWHM of both clumps in the deprojected image (FWHM$_{\rm intrinsic}$; \S\ref{sec:results_clumps}) are $\sim$4 times the local disk scale height $h$. Probably the particles traced by the sub-mm emission from the clumps have St$<$1, as particles with St$\sim$1 are expected to have drifted to the center of the clumps at the age of the system (smaller particles drift slower in both the radial and azimuthal directions, and are also more easily diffused by gas turbulence). Also, dust feedback onto gas tends to prevent particles from concentrating towards a single point and is expected when the local dust-to-gas ratio approaches unity \citep[e.g.,][]{fu14, miranda17}. We also rule out the presence  of multiply clustered sub-clumps separated azimuthally by more  than the beam size (6.5 AU). Such ``multiple clumps'' may be produced by dust trapping in vortices generated at a dead-zone edge due to viscosity transitions \citep[Figs. 6 \& 7]{miranda17}. Multiple clumps azimuthally separated by less than the beam size are still possible.

If the sub-mm continuum emission is spatially resolved and the dust is optically thin, the azimuthal peak-to-trough ratio at the position of a clump provides an estimate of the underlying contrast in the surface density of $\sim$mm-sized dust. However, since both clumps may only be marginally optically thin at their peaks (\S\ref{sec:results_clumps}), the measured azimuthal contrasts (4.4 for the south and 10 for the north clump) only provide a lower limit on the degree of dust concentration in the clumps.

\subsection{The Hypothetical Planets in the MWC 758 Disk}\label{sec:planet}

The central cavity of MWC 758 inward of $r \approx 51$ AU
may be opened by one or more planets inside \citep[e.g.,][]{zhu11, duffell15dong, dong15gap}. This cavity, however, has not been revealed in NIR scattered light down to an inner working angle of $\sim$0\farcs1 (16 AU; \citealt{reggiani18}; cf.~\citealt{benisty15} who reported a slight reduction in scattered light intensity inside the mm-wave cavity). \citet{dong12cavity} assign MWC 758 to the group of transitional disks with ``missing cavities.'' The data taken collectively are consistent with the gas inside the cavity having been depleted by no more than a factor of 10--100; mm-sized particles may be trapped at the overpressured cavity edge, but smaller micron-sized dust can still leak in with the gas \citep{dong12cavity}. The ellipticity of the cavity edge, and the possibility that the south clump may be a vortex formed at the cavity edge as triggered by the Rossby wave instability (see below), further constrain the properties of the disk and embedded planets (\S\ref{sec:cavity}). Determining the degree of gas depletion inside the cavity using gas emission observations \citep[e.g.,][]{vandermarel15, vandermarel16} will be essential.

The two spirals may be excited by a $\sim$5--10 $\mj$ planet exterior to the spirals at $\sim$100 AU, as proposed by \citet{dong15spiralarm}. A more massive companion even further away may also be able to drive the arms \citep[e.g.,][]{dong16hd100453}, but has been ruled out by observations \citep{reggiani18}. Massive, distant planets like the one predicted around MWC 758 have been found around other A-type stars (e.g., HR 8799 b, \citealt{marois08}; see also \citealt{bowler16}; note that MWC 758 will probably become a main sequence A-type star in the future). The formation of such planets is currently being investigated; they might be the outcomes of gravitational instability and disk fragmentation at early times \citep[e.g.,][]{kratter10, rafikov05}. The new ALMA data is consistent with the \citet{dong15spiralarm} hypothesis, as the presence of Arm~1 in sub-mm continuum emission suggests that it is mainly an overdensity rather than a shadow or a temperature structure. The fact that Arm~2 is not clearly revealed by ALMA may be caused by different dust-trapping capacities in the primary and secondary arms (\S\ref{sec:arm2}).

\begin{figure}
\begin{center}
\includegraphics[trim=0 0 0 0, clip,width=\textwidth,angle=0]{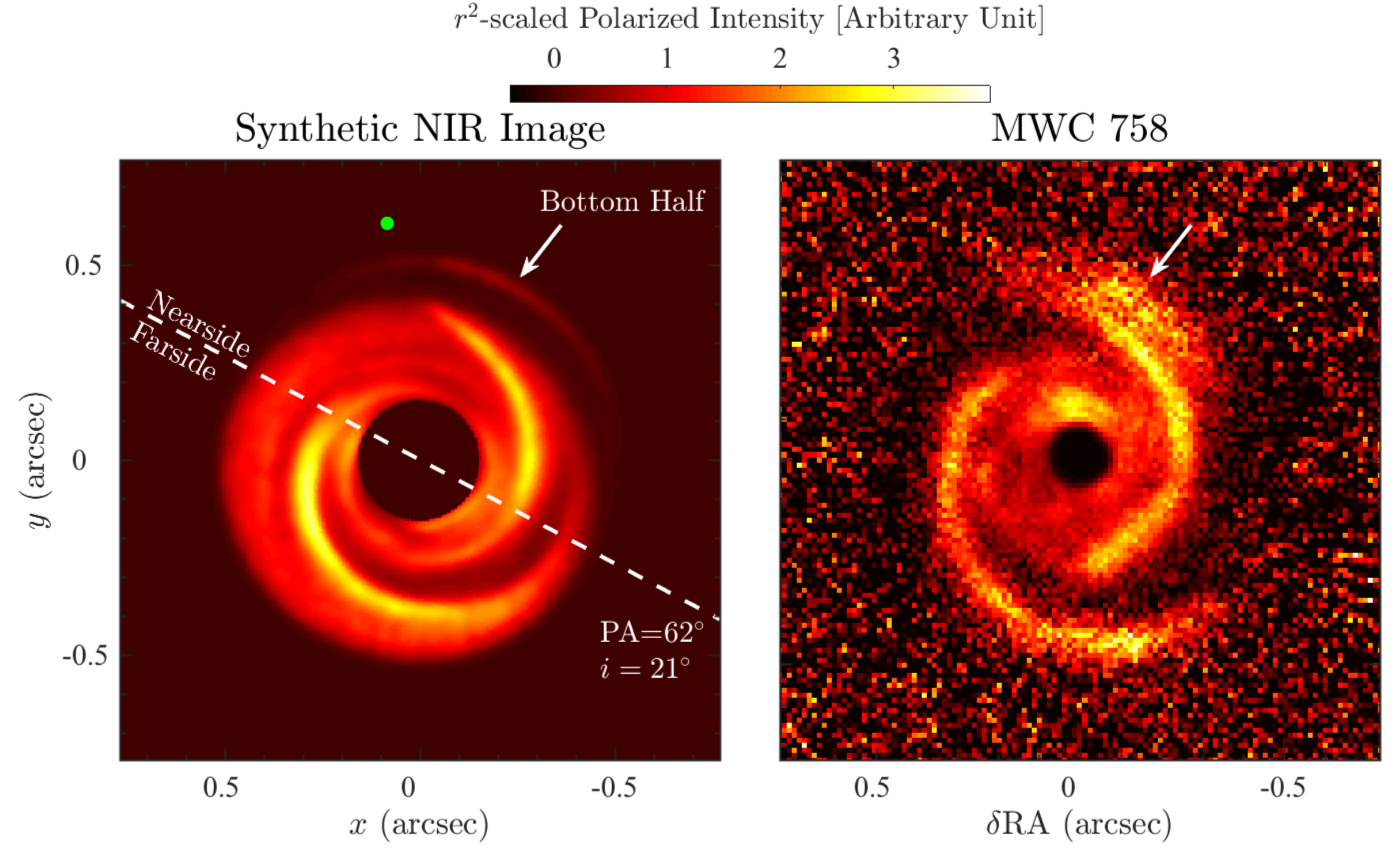}
\end{center}
\figcaption{{\it Left:} A simulated NIR polarized light image of a disk perturbed by a planet. {\it Right:} The $Y$-band (1 $\micron$) polarized light image of MWC 758 \citep{benisty15}. Both the model and the observation have been $r^2$-scaled. The simulation is based on the Model 6ISO125 in \citet{dong15spiralarm}, in which the planet has $\mplanet=6\times10^{-3}M_\star$ ($M_\star$ is the stellar mass) and is located at 100 AU. The disk is truncated at 80 AU (i.e., roughly at the inner edge of the gap opened by the planet). The synthetic observation is produced using the geometry of MWC 758 \citep{isella10, boehler18} --- the southeast side is the far side, inclination $i=21^\circ$, and position angle PA=62$^\circ$ (major axis marked by the dashed line). The green dot marks the projected location of the planet. If the outer disk is effectively truncated by the planet, the feature right next to the northern arm in the actual observation (indicated by the arrow) can be explained as the nearside edge of the bottom (obscured) half of the disk. See \S\ref{sec:planet} for discussion.
\label{fig:nir}}
\end{figure}

A massive (multi-$\mj$) companion at around 100 AU exterior to
the arms is also expected to truncate the disk interior to its 
orbit. Disk truncation is seen not just in scattered light but
also in ALMA observations of 
C$^{18}$O, revealing an outer gas disk edge at $\sim$100 AU 
\citep{boehler18}. We note that emission from $^{13}$CO, a more abundant species, is detected out to $\sim$150 AU \citep{boehler18}. 
Future gas emission modeling work is needed to quantify the spatial distribution of gas outside the main sub-mm disk ring.
Figure~\ref{fig:nir} shows a synthetic scattered light image 
based on the \citet{dong15spiralarm} disk model. It is identical 
to the original one (see Fig.~4 in that paper), except (1) the 
disk at $r \geq 0.8\rp$ (see, e.g., Fig.~2 of \citealt{fung16}) 
is removed prior to image synthesizing, and (2) the synthetic 
image is produced assuming not the face-on but the actual viewing geometry of the disk: PA=62$^\circ$, $i$=21$^\circ$, and with the northwest side being the side nearest the observer \citep{isella10, boehler18}. We note that the 3D hydro simulation has been run for 20 orbits, which is long enough for the spiral arms to be fully established and reach steady state, but not enough for the gap around the planet's orbit to be fully opened. To fully truncate the disk, simulations would need to be run for much longer time, prohibitively expensive in this case. Hence we truncate the disk manually. Future simulations run to the system age are needed to examine whether the truncation of the disk with the right radius and depletion can be naturally achieved by the planet.

Figure~\ref{fig:nir} shows that the nearside edge of the bottom half of the disk in the model image (left panel; indicated by the arrow), just outside and parallel to the tip of Arm~2, is clearly visible, and in good agreement with the location and orientation of the corresponding structure in the actual SPHERE image (right panel; indicated by the arrow). Note that while non-planet-based mechanisms can also produce a sharp outer edge in the distribution of $\sim$mm-sized dust due to dust radial drift \citep[e.g.,][]{birnstiel14}, the presence of this NIR feature requires a truncation and a sharp edge in the distribution of $\sim$$\micron$-sized small dust, usually well-coupled with the gas. Such a structure has also been seen in the \citet[Fig. 5]{dong16armviewing} study of spiral arm morphology in inclined disks, and in HD 100453 \citep{benisty17}, a disk with a truncated edge and two spirals. Our model feature is fainter than in the SPHERE image. This may be due to the specific properties of (sub-)$\micron$-sized dust in MWC 758 responsible for NIR scattered light (interstellar medium dust is assumed in the model; \citealt{kim94}). 

A few direct imaging campaigns have been carried out to look for this companion, with no confirmed detection at the moment. The achieved detection limit on the planet mass assuming ``hot start'' planet formation models, which predict a higher luminosity for a planet of a given mass at a given age than the ``warm'' or ``cold'' formation models \citep{spiegel12}, is approaching the theoretical prediction ($\sim5\mj$ at $r\sim$100 AU; \citealt{reggiani18}). The implication of such results is far from certain. Possibilities include (1) planets are not as luminous as the hot start model predictions, and (2) the system is older than the often assumed 3.5$\pm$2 Myr \citep{meeus12}, making the hypothetical planet colder and less luminous (particularly in hot start models). Age estimates for pre-main-sequence stars are usually made by placing the stars on the H-R diagram and comparing their locations with theoretical isochrones. \citet{meeus12} assumed the old Hipparcos distance of 279 pc for MWC 758 \citep{vanleeuwen07}, which is significantly greater than the new GAIA distance of 160 pc, and therefore overestimated its luminosity by a factor of 3. Re-examination of the stellar age using the GAIA distance is needed. Similarly, previous stellar mass ($M_\star$) determinations from isochrone  fitting are likely to be inaccurate. Disk kinematic studies can independently and more accurately determine $M_\star$ \citep[e.g.,][]{isella10, czekala15}, and thereby better
calibrate the hydrodynamic and radiative transfer simulations
of planet-disk interactions that all scale with $M_\star$.

The non-detection of the CPD in our ALMA observations also constrains CPD models. \citet{wu17} searched for but failed to detect mm continuum emission from the CPDs around five known planetary mass companions at large distances in other systems using ALMA (see also the non-detections by \citealt{bowler15} and \citealt{macgregor17}). The authors proposed that CPDs may be compact, having sizes on the order of 100 Jupiter radii or smaller, and optically thick, under which conditions the $\sim$mm flux density from such a disk is expected to be on the order of a $\mu$Jy, well below our rms noise level. Alternatively, the predicted companion outside the disk edge in MWC 758 may not have a CPD --- its CPD may have been lost through accretion, and it is not being replenished by an outer disk beyond the planet's orbit.

The two azimuthal clumps in $\sim$mm emission have also been proposed to be dust-trapping vortices generated by the Rossby wave instability (RWI) at the edges of planet-opened gaps \citep{lovelace99, li00}. The north clump has been shown to be more compact at longer wavelengths \citep{marino15mwc758}, consistent with being an azimuthal dust trap \citep[e.g.,][]{lyra13, zhu14stone}. Our new data showed that the radial extent (FWHM) of the dust distribution inside the clumps (particularly the north clump) may be $\sim$4$\times$ the local disk scale height, 
indicating the dust back-reaction onto the gas may be important (section \ref{sec:clump}). The proposed spiral-arm-driving planet exterior to the spirals may naturally sharpen the edge of the disk at $\sim$100 AU, triggering the RWI and subsequently the formation of a vortex as the north clump. 

Finally, the triple-ring structure best seen on the west side of the broad outer disk may also hint at the presence of planets. \citet[see also \citealt{bae17}]{dong17doublegap} showed that such a structure may be produced by a single planet in low-viscosity gas. The rich dynamical environment in MWC 758 with possible eccentric-cavity-opening planets inside and a spiral-driving planet outside makes simple model-data comparison difficult. The eccentricity of the observed gaps and rings ($e$$\approx$0.1) also adds an additional complication. Object-specific modeling taking into account relevant dynamical processes, including planet-planet interactions, is needed to comment on the origin of these narrow gaps.

\subsection{New Questions and Future Work}\label{sec:future}

The new data and analysis motivate future theoretical and observational studies:
\begin{enumerate}
\item \label{future:hydro}Gas+dust simulations to study whether and how planet-induced multiple spiral arms can trap dust particles, and the difference between the primary and secondary arm.
\item \label{future:modeling} Disk modeling to decompose the emission in the broad ring into components to examine whether unidentified segments of the two arms, particularly Arm~2, are hidden under other more prominent structures. 
\item Combining the above studies to examine which one of the two arms is the primary (companion-pointing) arm, should they be excited by a single companion. The outcome can direct future direct imaging campaigns to search for this companion.
\item Continuum emission observations at multiple sub-mm to cm wavelengths and spectral index analysis to examine the composition and size distribution of the dust in the two clumps. Comparing data with vortex dust trapping simulations to understand the effect of dust feedback onto the gas. 
\item Simulations to explore whether a companion at 100 AU or further can excite the two spirals, truncate the disk, and trigger the formation of the north clump at the truncated disk edge simultaneously.
\end{enumerate}


\section{Summary}\label{sec:summary}

We present Cycle 5 ALMA continuum emission observations of the protoplanetary disk around MWC 758 at 0.87~mm with a beam size of 43$\times$39 mas (6.8$\times$6.2 AU) and a rms noise level of 20 $\mu$Jy beam$^{-1}$. This system joins a few other protoplanetary disks revealed by high sensitivity ALMA observations with sub-$0\farcs05$ resolution (HL Tau, \citealt{brogan15}, \citealt{akiyama16hltau}; TW Hya, \citealt{andrews16}, \citealt{tsukagoshi16}, \citealt{huang18}; V883, \citealt{cieza16}; and V1247 Orionis, \citealt{kraus17}; see also GY 91, \citealt{sheehan18}). The high resolution sub-mm continuum map (Figure~\ref{fig:image}) reveals a central cavity $\sim$0\farcs32 (51 AU) in size, a broad outer disk extending to $\sim$0\farcs64 (102 AU) that can be decomposed into three faint rings, a south clump on the outer edge of the cavity, a north clump on the outer edge of the outer disk, a central emission point source around the star, and a spiral arm in the south. Basic measurements of the features are listed in Table~\ref{tab:measurement}. We compare the ALMA dataset with observations at other wavelengths, and propose possible interpretations for the observed disk structures. Our main conclusions are:
\begin{enumerate}
\item One of the two spiral arms imaged in NIR scattered light (Arm~1 in Figure~\ref{fig:image}$c$) is revealed in sub-mm continuum emission. This is the first time the sub-mm continuum counterpart of an NIR spiral arm is spatially resolved. The ALMA arm extends about 90$^\circ$ in the azimuthal direction, and 2/3 of its extent is located at a slightly larger (up to 8\%) stellocentric distance than its scattered light image counterpart. This offset can be explained as the surface (probed by NIR scattered light) and the midplane (probed by sub-mm continuum emission) of the same structure being projected to different locations on the sky (Figures~\ref{fig:schematic} and \ref{fig:simulation}). The azimuthal peak-to-trough contrast of the arm, 3.6, suggests it is mainly a density (emission optical depth) feature.

\item Contrary to Arm~1, Arm~2 (Figure~\ref{fig:image}$c$) is not clearly revealed in the ALMA image. We submit two possible explanations --- (A): Arm~2 does not trap $\sim$mm-sized dust as effectively as Arm~1 (due to, for example, the difference between the primary and secondary arms' gas perturbation amplitudes, as excited by one planet), making Arm~2 weaker / absent in sub-mm continuum emission; (B): Arm~2 is present at a slightly larger stellocentric distance coinciding with the east side of the inner ring and the clumps, making it difficult to identify. We favor the former explanation.

\item The cavity (and the inner ring which defines its edge) is not a circle centered on the star. This is not a projection effect. Once deprojected, the inner ring can be fit by an ellipse with eccentricity $e \approx 0.1$ and one focus on the star (Figure~\ref{fig:deproj}$c,d$). If fit by a circle, its center is at $\sim$30 mas from the GAIA stellar location, which roughly coincides with the detected central point source. This difference is $\sim$one order of magnitude larger than the expected ALMA astrometry uncertainty. The middle and outer rings also tentatively show the same eccentricity and orientation as the inner ring. Hydro simulations have shown that companions may open eccentric gaps under certain conditions \citep[e.g.,][]{kley06}.

\item Both the north ($r_{\rm deprojected}$=0\farcs53=85~AU) and the south clump ($r_{\rm deprojected}$=0\farcs31=50~AU) are spatially resolved in the radial and azimuthal directions. Their radial FWHM-to-radius ratios are $\sim$21\% (south) and 29\% (north), both being $\sim$4$\times$ the local disk aspect ratio $h/r$. Azimuthally their FWHMs are about 1/6 of a circle. The two clumps may be only marginally optically thin. 
Their significant radial widths suggest that emitting particles have
Stokes numbers smaller than unity and/or that they have been smeared out by dust-to-gas backreaction, as particles with St$\sim$1 are expected to have drifted to the center of the clumps at the age of the system.

\item The northwest feature parallel to Arm 2 in NIR imaging observations of the disk can be explained as the nearside edge of the bottom half of the disk, if the disk is truncated at the outer edge of the spiral arms (Figure~\ref{fig:nir}). The disk truncation, the two spiral arms, and the north clump all point toward an unseen external companion at around $\sim$100 AU.
\end{enumerate}


\section*{Acknowledgments}

We are grateful to Hui Li for useful discussions, Myriam Benisty for making available the SPHERE image of MWC 758, and an anonymous referee for constructive suggestions that largely improved the quality of the paper. Numerical calculations were performed on the SAVIO cluster provided by the Berkeley Research Computing program, supported by the UC Berkeley Vice Chancellor for Research and the Berkeley Center for Integrative Planetary Science. YP was  supported  by  the  Russian Science  Foundation grant  17-12-01168. This paper makes use of the following ALMA data: 2017.1.00492.S. ALMA is a partnership of ESO (representing its member states), NSF (USA), and NINS (Japan), together with NRC (Canada), MOST and ASIAA (Taiwan), and KASI (Republic of Korea), in cooperation with the Republic of Chile. The Joint ALMA Observatory is operated by ESO, AUI/NRAO, and NAOJ. The National Radio Astronomy Observatory is a facility of the National Science Foundation operated under cooperative agreement by Associated Universities, Inc. Y.H. is currently supported by the Jet Propulsion Laboratory, California Institute of Technology, under a contract with the National Aeronautics and Space Administration.


\bibliography{ms.bbl}

\begin{table}[]
\centering
\footnotesize
\resizebox{\textwidth}{!}{%
\begin{tabular}{@{}rccccccc@{}}
\toprule
Feature & Radius & PA & FWHM & FWHM & $I_{\rm peak}$ & $T_{\rm B,peak}$ & $I_{\rm peak}/I_{\rm trough}$ \\
 &  &  & Radial  & Azimuthal & mJy/beam & K &  \\
(1) & (2) & (3) & (4) & (5) & (6) & (7) & (8) \\ \midrule
Arm~1 & 0\farcs453$\pm$0\farcs006 & 175$^\circ$$\pm$1$^\circ$ & \dots & \dots & 0.51$\pm$0.02 & 9.0$\pm$0.2 & 3.6$\pm$0.5 \\
South Clump & 0\farcs306$\pm$0\farcs003 & 206$^\circ$$\pm$2$^\circ$ & 0\farcs075$\pm$0\farcs001 & 56$^\circ$$\pm$1$^\circ$ & 1.60$\pm$0.02 & 16.8$\pm$0.1 & 4.4$\pm$0.3 \\
North Clump & 0\farcs533$\pm$0\farcs006 & 335$^\circ$$\pm$1$^\circ$ & 0\farcs160$\pm$0\farcs003 & 54$^\circ$$\pm$1$^\circ$ & 1.28$\pm$0.02 & 14.7$\pm$0.1 & 10$\pm$1 \\
Inner Ring & 0\farcs319$\pm$0\farcs002 & \dots & 0\farcs085$\pm$0\farcs005 & \dots & 0.70$\pm$0.02 & 10.5$\pm$0.2 & \dots \\
Middle Ring & $\sim$0\farcs43 & \dots & \dots & \dots & 0.32$\pm$0.02 & 7.4$\pm$0.2 & \dots \\
Outer Ring & $\sim$0\farcs54 & \dots & \dots & \dots & 0.29$\pm$0.02 & 7.1$\pm$0.2 & \dots \\ \bottomrule
\end{tabular}
}
\caption{
Feature properties and their 1-$\sigma$ uncertainties measured in the deprojected ALMA 0.87 mm continuum emission map (Figure~\ref{fig:deproj}). Only quantities that can be robustly measured are listed. 
Column (1): Name of the feature (labeled in Figure~\ref{fig:image}$b$).
Column (2): Stellocentric radius at the peak. For the inner ring it is the semi-major axis in the ellipse fit in \S\ref{sec:results_cavity} (see also Figure~\ref{fig:deproj}$c$); for the middle and outer rings it is the semi-major axis in the tentative ellipse fits (these two rings are very incomplete).
Column (3): Position angle at the peak.
Column (4): Full width at half maximum in the radial direction. For the two clumps the measurement is done at the peak PA listed in (3). For the inner ring it is the average FWHM measured in the ``Ring West'' (0\farcs08) and ``Ring East'' regions (0\farcs09) in Figure~\ref{fig:deproj}($c$). The middle and outer rings do not permit such a measurement.
Column (5): Full width at half maximum in the azimuthal direction for the two clumps measured at the peak radius listed in (2).
Column (6): Peak intensity (surface brightness). For the inner ring it is the peak intensity on the east side, as the west side may contain contributions from the south clump. For the middle and outer rings it is the peak azimuthally-averaged intensity in the ``Ring West'' region in Figure~\ref{fig:deproj}($c$).
Column (7): Brightness temperature corresponding to the peak intensity in (6), converted using the Planck function.
Column (8): Azimuthal peak-to-trough intensity ratio. For the north clump and the spiral the trough intensity is taken as the minimum intensity along the circle at the radius listed in (2). For the south clump the trough intensity is taken as the minimum intensity along the elliptical inner ring.
\label{tab:measurement}}
\end{table}

\clearpage

\end{document}